\documentclass[aps,prc,preprint,amsmath,amssymb,showpacs,preprintnumbers,superscriptaddress]{revtex4-1}
\usepackage{CJK}
\usepackage{graphicx}
\usepackage{dcolumn}
\usepackage{bm}
\usepackage{mathrsfs}   
\usepackage{color,xcolor}  
\usepackage{multirow}  
\usepackage{booktabs} 

\usepackage{hyperref}

\newcommand{\ra}{\rangle}

\newcommand{\bbm}{\begin{bmatrix}}
\newcommand{\ebm}{\end{bmatrix}}
\newcommand{\bBm}{\begin{Bmatrix}}
\newcommand{\eBm}{\end{Bmatrix}}
\newcommand{\bpm}{\begin{pmatrix}}
\newcommand{\epm}{\end{pmatrix}}

\begin{document}


\title{Properties of $^{208}$Pb predicted from the relativistic equation of state in the full Dirac space}

\author{Hui Tong}
\affiliation{Department of Physics, Chongqing University, Chongqing 401331, China}
\affiliation{Strangeness Nuclear Physics Laboratory, RIKEN Nishina Center, Wako, 351-0198, Japan}

\author{Jing Gao}
\affiliation{School of Physics, Nankai University, Tianjin 300071, China}

\author{Chencan Wang}
\affiliation{School of Physics and Astronomy, Sun-Sat-Sen 
University, Zhuhai 519082, China}

\author{Sibo Wang}
\email{sbwang@cqu.edu.cn}
\affiliation{Department of Physics, Chongqing University, Chongqing 401331, China}



\date{\today}

\begin{abstract}

Relativistic Brueckner-Hartree-Fock (RBHF) theory in the full Dirac space allows one to determine uniquely the momentum dependence of scalar and vector components of the single-particle potentials.
In order to extend this new method from nuclear matter to finite nuclei, as a first step, properties of $^{208}$Pb are explored by using the microscopic equation of state for asymmetric nuclear matter and a liquid droplet model.
The neutron and proton density distributions, the binding energies, the neutron and proton radii, and the neutron skin thickness in $^{208}$Pb are calculated.
In order to further compare the charge densities predicted from the RBHF theory in the full Dirac space with the experimental charge densities, the differential cross sections and the electric charge form factors in the elastic electron-nucleus scattering are obtained by using the phase-shift analysis method.
The results from the RBHF theory are in good agreement with the experimental data.
In addition, the uncertainty arising from variations of the surface term parameter $f_0$ in the liquid droplet model is also discussed.

\end{abstract}



\maketitle


\section{Introduction}\label{SecI}

Exploring the equation of state (EOS) of neutron rich matter has attracted considerable attentions in the field of nuclear physics and astrophysics, particularly the poorly determined density dependence of the symmetry
energy~\cite{Lattimer_2000_PR333_121,Li_2008_PR464_113,Oertel2017RMP_89_015007,Burgio2021PPNP_120_103879}.
In terrestrial laboratories, the thickness of the neutron skin in neutron-rich atomic nuclei, e.g., $^{208}$Pb, has been identified as an ideal laboratory observable to constrain the EOS of neutron rich matter during the last decades~\cite{ABBrown_2000PRL-85-5296,Centelles_2009_PRL102_122502,Roca-Maza_2011-PRL-106-252501,Reed2021PRL}.
The neutron skin thickness is defined as the difference between the neutron ($r_n$) and proton ($r_p$) root-mean-square (rms) radii:~$\Delta r_{np}=r_n-r_p$.
Currently, precise data on $r_p$ are available.
In contrast, as neutrons are uncharged, the determination on $r_n$ still suffer from uncontrolled uncertainties due to hadron dynamics which requires model assumptions to deal with the strong force~\cite{Thiel2019JPG}. A model-independent method to probe the neutron densities was also proposed, i.e., parity-violating electron scattering~\cite{Donnelly1989NPA,Horowitz2001PRC}. Recently, an improved value for the neutron skin thickness of $^{208}$Pb has been reported by the updated Lead Radius EXperiment (PREX-II):~$\Delta r_{np}=0.283\pm 0.071$~fm~\cite{Adhikari2021PRL}.
By using the strong correlation between $\Delta r_{np}$ and the slope of the symmetry energy $L$ calculated from several sets of relativistic energy density functionals, a new value of $L=106\pm37$~MeV was reported and it is larger than those from other theoretical and experimental estimations~\cite{Reed2021PRL}.
This result provides a challenge to our present understanding of the nuclear matter EOS, particularly the density dependence of symmetry energy.



In theoretical studies, the predictions from the density functionals span a fairly wide range of neutron skin thickness for $^{208}$Pb, since the isovector channels are loosely constrained in the fitting procedures~\cite{Roca-Maza2011_PRL-106-252501}. For \emph{ab initio} methods, calculating such a heavy nucleus directly is difficult due to the huge computational cost.
Recently, $\Delta r_{np}=0.14-0.20$~fm for $^{208}$Pb was predicted by the nonrelativistic \emph{ab initio} calculations from the chiral effective field theory (EFT) with realistic two- and three-nucleon forces~\cite{Hu2022Nature}.
In the relativistic framework, Relativistic Brueckner-Hartree-Fock (RBHF) theory is one
of the most successful \emph{ab initio} theories based on realistic two-nucleon forces only~\cite{Brockmann1990_PRC42-1965}.
Comparing to the nonrelativistic Brueckner-Hartree-Fock (BHF) theory with two-nucleon forces only, the RBHF theory improves the description considerably, both for the nuclear matter and finite nuclei~\cite{SHEN-SH2017_PRC96-014316,SHEN-SH2019_PPNP109-103713}. Due to considerable numerical difficulties, the investigation of $^{208}$Pb within a full solution in the relativistic framework is not available.     
The present fully self-consistent RBHF theory has been accomplished only for the neutron skin thickness in the medium-mass nucleus $^{48}$Ca~\cite{SHEN-SH2018_PRC97-054312}.
To extend the RBHF theory from the nuclear matter to finite nuclei in previous studies, most investigations were based on the local density approximation (LDA)~\cite{Negele1970PRC} and the liquid droplet model~\cite{OYAMATSU1998NPA}.
LDA aims to find an effective interaction which can reproduce the nuclear matter properties from the RBHF theory, while the key point of the liquid droplet model is to combine an energy functional based on the semiempirical mass
formula with the EOS from RBHF theory~\cite{Alonso2003PRC,Sammarruca2009_PRC79_057301,Sammarruca2016PRC,Sammarruca2016_PRC94_044311}.

One of the most essential ingredient in the RBHF calculations for nuclear matter properties is to identify the single-particle potentials of the nucleons.
However, to avoid the complexity of extracting self-consistently the scalar and vector components of the single-particle potentials from the $G$ matrix, the momentum-independence approximation method and the projection method are usually introduced in the Dirac space with positive-energy states (PESs) only~\cite{Brockmann1990_PRC42-1965,Gross-Boelting1999_NPA648-105,Schiller2001_EPJA11-15}.
These two frequently used approximations lead to contradictory results for the isospin dependence of the Dirac mass~\cite{Ulrych1997_PRC56-1788}.
Therefore, the mapping from nuclear matter to finite nuclei is far from unique because of the uncertainties of the RBHF theory in nuclear matter~\cite{Brockmann1992_PRL68-3408,Fritz1993_PRL71_46,Fuchs1995_PRC52_3043,Ulrych1997_PRC56-1788,SHEN-H1997_PRC55-1211,Ma2002_PRC66_024321,vanDalen2011_PRC84_024320}.

Recently, by considering the PESs and negative-energy states (NESs) simultaneously, a self-consistent RBHF calculation is developed in the full Dirac space~\cite{WANG-SB2021_PRC103-054319,2022-Wang-SIBO-PhysRevC.106.L021305}.
In this case, the momentum dependence for the scalar and vector components of the single-particle potentials are  determined uniquely, since the matrix elements of single-particle potential operator can be decomposed in the full Dirac space.
The RBHF theory in the full Dirac space has been successfully applied to study the symmetric nuclear matter (SNM)~\cite{WANG-SB2021_PRC103-054319,WANG-SB2022_PhysRevC.105.054309}, asymmetric nuclear matter (ANM)~\cite{2022-Wang-SIBO-PhysRevC.106.L021305,Tong_2022-AstrophysicsJ930.137} and neutron star properties~\cite{Tong_2022-AstrophysicsJ930.137,Wang2022_PRC106_045804}.
Especially, this new method has clarified the long-standing controversy about the isospin dependence of the Dirac mass in the RBHF calculations of ANM~\cite{2022-Wang-SIBO-PhysRevC.106.L021305}.

With the success of RBHF theory in the full Dirac space in nuclear matter, it is interesting and timely to extend this method to study finite nuclei.
As a first step, the EOS from the RBHF theory in the full Dirac space and a liquid droplet
model are used to study the properties of $^{208}$Pb.
To further compare the charge densities predicted from the RBHF theory in the full Dirac space with the experimental charge densities, the differential cross sections (DCS) and the electric charge form factors in the elastic electron-nucleus scattering are discussed by using the phase-shift analysis method.

In Sec.~\ref{SecII}, the theoretical framework of the RBHF theory in the full Dirac space, the liquid droplet
model, and the phase-shift analysis method are introduced. The calculated results and discussions are presented in Sec.~\ref{SecIII}.
Finally, a summary is given in Sec.~\ref{SecIV}.


\section{Theoretical framework} \label{SecII}

\subsection{Relativistic Brueckner-Hartree-Fock theory in the full Dirac space}

To describe the single-particle motion of a nucleon in nuclear matter from the RBHF theory, the essential point is to use the Dirac equation,
\begin{equation}\label{DiracEquation}
  \left\{ \bm{\alpha}\cdot\bm{p}+\beta \left[M+\mathcal{U}_\tau(\bm{p})\right] \right\} u_\tau(\bm{p},s)
  = E_{\bm{p},\tau}u_{\tau}(\bm{p},s), \quad \tau = n,p,
\end{equation}
where $\bm{\alpha}$ and $\beta$ are the Dirac matrices, $M$ is the nucleon mass, $\bm{p}$ and $E_{\bm{p},\tau}$ are the momentum and single-particle energy, $s$ and $\tau$ are the spin and isospin.
The single-particle potential $\mathcal{U}_\tau$ in the infinite, uniform nuclear matter can be decomposed in its Lorentz form~\cite{Serot1986_ANP16-1}:
\begin{equation}\label{SPP}
  \mathcal{U}_\tau(\bm{p}) = U_{S,\tau}(p)+ \gamma^0U_{0,\tau}(p) + \bm{\gamma\cdot\hat{p}}U_{V,\tau}(p),
\end{equation}
where $U_{S,\tau}$ is the scalar potential, $U_{0,\tau}$ and $U_{V,\tau}$ are the timelike and spacelike parts of the vector potential.
$\hat{\bm{p}}=\bm{p}/|\bm{p}|$ is the unit vector parallel to the momentum $\bm{p}$.
By introducing the following effective quantities, $\bm{p}^*_\tau=\bm{p}+\hat{\bm{p}}U_{V,\tau}(p)$, $M^*_{\bm{p},\tau}=M+U_{S,\tau}(p)$, and $E^*_{\bm{p},\tau}=E_{\bm{p},\tau}-U_{0,\tau}(p)$,
the solutions of Eq.~\eqref{DiracEquation} in the full Dirac space are
\begin{equation}\label{DiracSpinor}
  u_\tau(\bm{p},s) =\ \sqrt{\frac{E_{\bm{p},\tau}^*+M_{\bm{p},\tau}^*}{2M_{\bm{p},\tau}^*}}
  		\bbm 1 \\ \frac{\bm{\sigma}\cdot\bm{p}^*_\tau}{E_{\bm{p},\tau}^*+M_{\bm{p},\tau}^*}\ebm \chi_s\chi_\tau,~~~v_\tau(\bm{p},s) =\ \gamma^5u_\tau(\bm{p},s),
\end{equation}
where $u_\tau$ and $v_\tau$ are in-medium baryon spinors with positive and negative energies, $\chi_s$ and $\chi_\tau$ are the spin and isospin doublets.


The baryon spinor can be calculated exactly once the scalar and vector components of the single-particle potentials are determined.
To this end, three matrix elements of the single-particle potential $\mathcal{U}_\tau$ in the full Dirac space are defined as~\cite{Anastasio1981_PRC23-2273,Poschenrieder-1988-PRC.38.471},
\begin{subequations}\label{Sigma++-+--}
  \begin{align}
    \Sigma^{++}_\tau(p)
    =&\   \bar{u}_\tau(\bm{p},1/2) \mathcal{U}_\tau(\bm{p}) u_\tau(\bm{p},1/2)
          = U_{S,\tau}(p) + \frac{E^*_{\bm{p},\tau}}{M^*_{\bm{p},\tau}} U_{0,\tau}(p)
          				  + \frac{p^*_\tau}{M^*_{\bm{p},\tau}} U_{V,\tau}(p), \label{Sigma++def}\\
    \Sigma^{-+}_\tau(p)
    =&\  \bar{v}_\tau(\bm{p},1/2) \mathcal{U}_\tau(\bm{p}) u_\tau(\bm{p},1/2)
          = \frac{p^*_\tau}{M^*_{\bm{p},\tau}} U_{0,\tau}(p)
          	+ \frac{E^*_{\bm{p},\tau}}{M^*_{\bm{p},\tau}} U_{V,\tau}(p), \label{Sigma-+def}\\
    \Sigma^{--}_\tau(p)
    =&\   \bar{v}_\tau(\bm{p},1/2) \mathcal{U}_\tau(\bm{p}) v_\tau(\bm{p},1/2)
          = - U_{S,\tau}(p) + \frac{E^*_{\bm{p},\tau}}{M^*_{\bm{p},\tau}} U_{0,\tau}(p)
          					+ \frac{p^*_\tau}{M^*_{\bm{p},\tau}} U_{V,\tau}(p) .\label{Sigma--def}
  \end{align}
\end{subequations}
When $\Sigma^{++}_\tau$, $\Sigma^{-+}_\tau$, and $\Sigma^{--}_\tau$ are determined, the single-particle potentials in Eq.~\eqref{SPP} can be obtained uniquely from
\begin{subequations}\label{Sigma2US0V}
  \begin{align}
    U_{S,\tau}(p) = &\ \frac{\Sigma^{++}_\tau(p)-\Sigma^{--}_\tau(p)}{2},\\
    U_{0,\tau}(p) = &\ \frac{E^*_{\bm{p},\tau}}{M^*_{\bm{p},\tau}}\frac{\Sigma^{++}_\tau(p)+\Sigma^{--}_\tau(p)}{2}
    					 - \frac{p^*_\tau}{M^*_{\bm{p},\tau}}\Sigma^{-+}_\tau(p),\\
    U_{V,\tau}(p) = &\ -\frac{p^*_\tau}{M^*_{\bm{p},\tau}}\frac{\Sigma^{++}_\tau(p)+\Sigma^{--}_\tau(p)}{2}
    				  + \frac{E^{*}_{\bm{p},\tau}}{M^*_{\bm{p},\tau}} \Sigma^{-+}_\tau(p).
  \end{align}
\end{subequations}
Therefore, the matrix elements for the positive-energy solutions, the elements coupling positive- with negative-energy solutions given in Eq.~\eqref{DiracSpinor} and those for the negative-energy solutions, i.e., $\Sigma^{++}_\tau$, $\Sigma^{-+}_\tau$, and $\Sigma^{--}_\tau$ should be calculated simultaneously.
These three matrix elements can be evaluated through the effective $NN$ interactions $G$ matrix in the full Dirac space
\begin{subequations}\label{Gm2Sigma}
  \begin{align}
    \Sigma^{++}_\tau(p) = &\ \sum_{s'\tau'} \int^{k^{\tau'}_F}_0 \frac{d^3p'}{(2\pi)^3}
    			\frac{M^*_{\bm{p}',\tau'}}{E^*_{\bm{p}',\tau'}}
    			\langle \bar{u}_\tau(\bm{p},1/2) \bar{u}_{\tau'}(\bm{p}',s')| \bar{G}^{++++}(W)|
    			u_\tau(\bm{p},1/2)u_{\tau'}(\bm{p}',s')\ra, \label{Sigma++} \\
    \Sigma^{-+}_\tau(p) = &\ \sum_{s'\tau'} \int^{k^{\tau'}_F}_0 \frac{d^3p'}{(2\pi)^3}
    			\frac{M^*_{\bm{p}',\tau'}}{E^*_{\bm{p}',\tau'}}
			    \langle \bar{v}_\tau(\bm{p},1/2) \bar{u}_{\tau'}(\bm{p}',s')| \bar{G}^{-+++}(W)|
			    u_\tau(\bm{p},1/2)u_{\tau'}(\bm{p}',s')\ra, \label{Sigma-+} \\
    \Sigma^{--}_\tau(p) = &\ \sum_{s'\tau'} \int^{k^{\tau'}_F}_0 \frac{d^3p'}{(2\pi)^3}
    			\frac{M^*_{\bm{p}',\tau'}}{E^*_{\bm{p}',\tau'}}
    			\langle \bar{v}_\tau(\bm{p},1/2) \bar{u}_{\tau'}(\bm{p}',s')| \bar{G}^{-+-+}(W)|
    			v_\tau(\bm{p},1/2)u_{\tau'}(\bm{p}',s')\ra, \label{Sigma--}
  \end{align}
\end{subequations}
where $k_F^{\tau'}$ specifies the Fermi momentum for nucleon $\tau'$. 
$\bar{G}$ is the antisymmetrized $G$ matrix, where $\pm$ in the superscript denotes PESs or NESs.

In the RBHF theory, the in-medium covariant Thompson equation~\cite{Brockmann1990_PRC42-1965} is one of the most widely used equations to derive the $G$ matrix in nuclear matter,
\begin{equation}\label{ThomEqu}
	\begin{split}
  G_{\tau\tau'}(\bm{q}',\bm{q}|\bm{P},W)
  =&\ V_{\tau\tau'}(\bm{q}',\bm{q}|\bm{P})
  + \int \frac{d^3k}{(2\pi)^3}
  V_{\tau\tau'}(\bm{q}',\bm{k}|\bm{P}) \\
    & \times \frac{M^{*}_{\bm{P}+\bm{k},\tau}M^{*}_{\bm{P}-\bm{k},\tau'}}{E^*_{\bm{P}+\bm{k},\tau}E^*_{ \bm{P}-\bm{k},\tau'}}
    \frac{Q_{\tau\tau'}(\bm{k},\bm{P})}{W-E_{\bm{P}+\bm{k},\tau}-E_{\bm{P}-\bm{k},\tau'}}  G_{\tau\tau'}(\bm{k},\bm{q}|\bm{P},W),
  \end{split}
\end{equation}
where $\tau\tau'=nn,~pp$ or $np$, $W$ is the starting energy, $V_{\tau\tau'}$ denotes a realistic bare $NN$ interaction and Bonn potentials~\cite{Machleidt1989_ANP19-189} are used here.
$\bm{P}$ is the center-of-mass momentum and $\bm{k}$ is the relative momentum of the two interacting nucleons.
$\bm{q}, \bm{k}$ and $\bm{q}'$ are the initial, intermediate, and final relative momenta of the two nucleons scattering in nuclear matter, respectively. $Q_{\tau\tau'}(\bm{k},\bm{P})$ is the Pauli operator,
\begin{equation}\label{PauliBlockOperator}
  Q_{\tau\tau'}(\bm{k},\bm{P})
  = \left\{
	\begin{array}{cl}
	1  &   \quad  |\bm{P}+\bm{k}|>k_F^{\tau}~
			\textrm{and}~|\bm{P}-\bm{k}|>k_F^{\tau'}, \\
	0  &   \quad  \textrm{otherwise},
	\end{array}\right.
\end{equation}
which prevents $NN$ scattering into occupied states in the nuclear medium.

Through equations~\eqref{DiracEquation}, \eqref{Sigma2US0V}, \eqref{Gm2Sigma}, and \eqref{ThomEqu}, the $G$ matrix is self-consistently calculated with the single-particle potentials in the standard RBHF iterative procedure.
When the iteration has converged, the binding energy per nucleon in nuclear matter can be calculated by
\begin{equation}\label{E/A}
  \begin{split}
  E/A
  =&\ \frac{1}{\rho} \sum_{s,\tau} \int^{k^\tau_F}_0 \frac{d^3p}{(2\pi)^3} \frac{M^*_{\bm{p},\tau}}{E^*_{\bm{p},\tau}}
  \langle \bar{u}_\tau(\bm{p},s)| \bm{\gamma}\cdot\bm{p} + M |u_\tau(\bm{p},s)\ra - M \\
    &\ + \frac{1}{2\rho} \sum_{s,s',\tau,\tau'} \int^{k^\tau_F}_0 \frac{d^3p}{(2\pi)^3} \int^{k^{\tau'}_F}_0
     	  \frac{d^3p'}{(2\pi)^3} \frac{M^*_{\bm{p},\tau}}{E^*_{\bm{p},\tau}}\frac{M^*_{\bm{p}',\tau'}}{E^*_{\bm{p}',\tau'}} \\
  	&\ \times \langle \bar{u}_\tau(\bm{p},s) \bar{u}_{\tau'}(\bm{p}',s') |\bar{G}^{++++}(W)| u_\tau(\bm{p},s) u_{\tau'}(\bm{p}',s') \ra.
  \end{split}
\end{equation}

\subsection{From nuclear matter to finite nuclei}

To connect the nuclear matter EOS to properties of finite nuclei, the energy of a nucleus is written in
terms of a volume, a surface, and a Coulomb term motivated by a liquid droplet model~\cite{OYAMATSU1998NPA,Alonso2003PRC},
\begin{equation}\label{equldm}
  \begin{split}
    e(Z,A)=&\int d^3r E/A(\rho_n(r),\rho_p(r))\rho(r)+f_0\int d^3r |\nabla\rho(r)|^2\\
    &+\frac{e^2}{4\pi\epsilon_0}(4\pi)^2\int_0^\infty dr' r' \rho_p(r')\int_0^{r'}dr r^2 \rho_p(r),
  \end{split}
\end{equation}
where the volume term is calculated directly from the EOS in the RBHF theory with the density $\rho(r) = \rho_n(r) + \rho_p(r)$.
The constant $f_0$ is the surface energy parameter.
The proton and neutron density distributions are parameterized as standard Thomas-Fermi distributions
\begin{equation}
  \rho_i(r)=\frac{a_i}{1+e^{(r-R_i)/d_i}}  \quad (i = n,~p),
\end{equation}
where the free parameters of radius $R_i$ and diffuseness $d_i$ are extracted by minimization of the energy, while $a_i$ is obtained by normalizing the proton (neutron) distribution to $Z(N)$,
\begin{equation}
  \int d^3r \rho_i(r)=
  \begin{cases}
    A-Z,&i=n,\\
    Z,&i=p.
  \end{cases}
\end{equation}
The neutron skin thickness is defined as
\begin{equation}
  \begin{split}
    \Delta r_{np} &= r_n - r_p,
  \end{split}
\end{equation}
where $r_n$ and $r_p$ are the rms radii of the neutron and proton density distributions.


\subsection{Phase-shift analysis method}

Phase-shift analysis method is used to calculate the DCS and the electric charge form factors for elastic electron scattering from the charge densities obtained by the RBHF theory in the full Dirac space.
To obtain the DCS of the elastic electron scattering, one needs to solve the Dirac equation
\begin{equation}
  [\bm{\alpha}\cdot \bm{p} +\beta m+ V(r)]\psi(\bm{r})=E\psi(\bm{r}),
\end{equation}
where $E$ and $\bm{p}$ are the energy and momentum of the incident electrons, $m$ is the rest mass of the electron, and $V(r)$ is the Coulomb potential between the electron and the nucleus.
Since $V(r)$ is assumed to be spherical~\cite{Bjorken1965RQM},
\begin{equation}
    V(r) = - {e^2\over 4\pi\epsilon_0} \int d^3r' {\rho_\mathrm{ch}(r')\over
    |\bm{r}-\bm{r}'|}, 
\end{equation}
with $\rho_\mathrm{ch}(r)$ being the charge distribution.
The wave function $\psi(\bm{r})$ can be expressed by a series of spherical spinors with definite angular momenta,
\begin{equation}
  \psi(\bm{r})=\frac{1}{r}
  \begin{bmatrix}
    P(r)\Omega_{\kappa,m_j}(\theta,\phi)\\
    iQ(r)\Omega_{-\kappa,m_j}(\theta,\phi)
  \end{bmatrix},
\end{equation}
where $\Omega$ are the spherical spinors, $\kappa = (l-j)(2j+1)$ is defined as the relativistic quantum number, $j$ and $l$ are the total and orbital angular momentum quantum numbers. 
The radial wave functions $P(r)$ and $Q(r)$ satisfy the following
coupled differential equations:
\begin{subequations}\label{eq:rwavecde}
  \begin{align}
    &\frac{dP}{dr}=-\frac{\kappa}{r}P(r)+[E-V(r)+2m]Q(r),\\
    &\frac{dQ}{dr}=-[E-V(r)]P(r)+\frac{\kappa}{r}Q(r),
  \end{align}
\end{subequations}
The spin-up $\delta_l^+$ and spin-down $\delta_l^-$ phase shifts for the partial wave with orbital angular momentum $l$ can be obtained by solving the coupled radial equations \eqref{eq:rwavecde} with the asymptotic behavior.
Then one can calculate the direct scattering amplitude
\begin{equation}
  f(\theta)=\frac{1}{2ik}\sum_{l=0}^{\infty}[(l+1)(e^{2i\delta_l^+}-1)+l(e^{2i\delta_l^-}-1)]P_l(\cos\theta)
\end{equation}
and the spin-flip scattering amplitude
\begin{equation}
  g(\theta)=\frac{1}{2ik}\sum_{l=0}^{\infty}[e^{2i\delta_l^-}-e^{2i\delta_l^+}]P_l^1(\cos\theta),
\end{equation}
where $k$ is the wave number of the projectile electron, $P_l(\cos\theta)$ and $P_l^1(\cos\theta)$ are Legendre polynomials and associated Legendre functions.
The DCS for elastic electron scattering off nucleus can be given by
\begin{equation}
  \frac{d\sigma}{d\Omega}=|f(\theta)|^2+|g(\theta)|^2.
\end{equation}

\section{Results and discussion}\label{SecIII}

\begin{table}[htbp]
  \centering
  \caption{
   Parameters of the density distributions for $^{208}$Pb calculated by the RBHF theory in the full Dirac space.
  }
  \begin{tabular}{ccccccccc}
    \hline\hline

    \multirow{2}{*}{Method} & \multirow{2}{*}{Potential} & $f_0$ & $a_n$ & $R_n$ & $d_n$ & $a_p$ & $R_p$ & $d_p$ \\
 & & (MeV$\cdot$fm$^{5}$) & (fm$^{-3}$) & (fm) & (fm) & (fm$^{-3}$) & (fm) & (fm) \\

\hline

\multirow{2}{*}{Full Dirac Space} & \multirow{2}{*}{Bonn A} & 60 & 0.096 & 6.62 & 0.60 & 0.065 & 6.55 & 0.56 \\
&  & 70 & 0.095 & 6.62 & 0.65 & 0.064 & 6.54 & 0.61 \\

    \hline\hline
  \end{tabular}
  \label{tab1}
\end{table}

The RBHF calculation is performed for the nuclear matter EOS in the full Dirac space, where the bare nucleon-nucleon ($NN$) interaction is adopted as relativistic potential Bonn A~\cite{Machleidt1989_ANP19-189}.
With potential Bonn A, The empirical saturation properties of SNM are well reproduced from the RBHF calculations in the full Dirac space ~\cite{WANG-SB2021_PRC103-054319}.
From the EOS calculated by the RBHF theory and the energy functional based on the semiempirical mass formula~\cite{OYAMATSU1998NPA,Alonso2003PRC}, the two-parameter Fermi functions for neutron and proton densities of $^{208}$Pb are obtained.
The parameters for $a_i$,~$R_i$,~$d_i$ are listed
in Table~\ref{tab1}, where $i=n,p$.
It should be noted that the constant $f_0$ in Eq.~\eqref{equldm} from the surface term is typically obtained by fitting to $\beta$-stable nuclei and found to be about 60--70 MeV$\cdot$fm$^5$~\cite{Oyamatsu2010PRC}.
In order to study how this uncertainty impacts the corresponding predictions for the properties of $^{208}$Pb, Table~\ref{tab1} shows the results obtained with $f_0$ = 60 MeV$\cdot$fm$^5$ and those with $f_0$ = 70 MeV$\cdot$fm$^5$.
It can be seen that these two values of $f_0$ lead to different parameters, especially for $d_i$.

\begin{figure}[htbp]
  \centering
  \includegraphics[width=12.0cm]{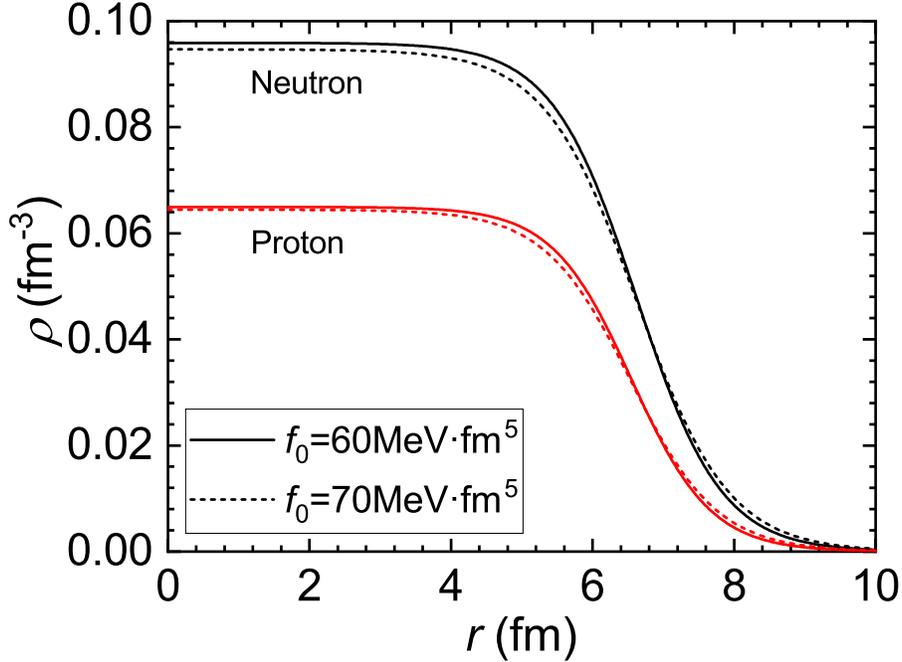}
  \caption{
  (Color online) Neutron and proton density distributions for $^{208}$Pb calculated by the RBHF theory in the full Dirac space with the potential Bonn A.
  The solid lines and dashed lines are obtained by using $f_0 = 60$ MeV$\cdot$fm$^5$ and $f_0 = 70$ MeV$\cdot$fm$^5$, respectively.
  }
  \label{Fig1}
\end{figure}

In Fig.~\ref{Fig1}, the radial dependence of the neutron and proton densities for $^{208}$Pb are calculated by the RBHF theory in the full Dirac space with the potential Bonn A.
The values of central densities from $f_0 = 60$ MeV$\cdot$fm$^5$ are larger than $f_0 = 70$ MeV$\cdot$fm$^5$ for both neutron and proton densities.
In contrast, the densities from $f_0 = 70$ MeV$\cdot$fm$^5$ are larger than those from $f_0 = 60$ MeV$\cdot$fm$^5$ when the radius becomes larger than $6.8$ fm.

\begin{table}[htbp]
  \centering
  \caption{
  Neutron radii, proton radii, and neutron skins calculated by the RBHF theory in the full Dirac space.
  }
  \begin{tabular}{ccccccccc}
    \hline\hline

    \multirow{2}{*}{Method} & \multirow{2}{*}{Potential} & $f_0$ & $r_n$ & $r_p$ & $\Delta r_{np}$ \\
 & &(MeV$\cdot$fm$^{5}$)& (fm) & (fm) &(fm) \\

\hline

\multirow{2}{*}{Full Dirac Space} & \multirow{2}{*}{Bonn A} &    60 & 5.59 & 5.48 & 0.11 \\
&  & 70 & 5.67 & 5.55 & 0.12 \\

    \hline\hline
  \end{tabular}
  \label{tab2}
\end{table}

From the density distributions obtained by the RBHF theory in the full Dirac space, the neutron radii $r_n$, proton radii $r_p$, and neutron skins $\Delta r_{np}$ for $^{208}$Pb can be extracted.
In Table~\ref{tab2}, the results for $r_n$, $r_p$, and $\Delta r_{np}$ are 5.59, 5.48, and 0.11 fm from $f_0=60$ MeV$\cdot$fm$^{5}$, as compared to 5.67, 5.55, and 0.12 fm from $f_0=70$ MeV$\cdot$fm$^{5}$.
One can see that the predictions from the latter one are even larger.
Recently, an accurate measurements of the neutron skin of $^{208}$Pb from
PREX-II is available~\cite{Adhikari2021PRL}.
Combined with the previous measurement~\cite{Abrahamyan2012PRL,Horowitz2012PRC}, the extracted neutron skin is $\Delta r_{np} = 0.283 \pm 0.071$ fm.
Furthermore, a value of the slope of the symmetry energy $L = (106\pm 37)$ MeV was reported through exploiting the strong correlation between the neutron skin of $^{208}$Pb and $L$ from a specific class of relativistic energy density functionals~\cite{Reed2021PRL}.
It should be noted that this result systematically overestimates current limits from both theoretical approaches and experimental measurements and suggests that the EOS at the typical densities found in atomic nuclei is stiff.
It can be also found that the neutron skin from PREX-2 is larger than those predicted by the RBHF theory in the full Dirac space.
This indicates that the EOS from the RBHF theory at the typical densities is rather soft.
In contrast, the neutron skin from the RBHF theory are close to other empirical values, such as $\Delta r_{np} = 0.15 \pm 0.08$ fm and $\Delta r_{np} = 0.14 \pm 0.10$ from pionic probes~\cite{Friedman2012NPA}, $\Delta r_{np} = 0.18 \pm 0.05$ fm~\cite{Garcia-Recio1992NPA_547_473} from the pionic atom potentials with varying radial parameters of the neutron distributions, and $\Delta r_{np} = 0.18 \pm 0.05$ fm~\cite{Tsang2012PRC} from symmetry energy constraints.
Therefore, the next generation of terrestrial experiments and astronomical observations are necessary to improve our understanding of the EOS.

\begin{figure}[htbp]
  \centering
  \includegraphics[width=12.0cm]{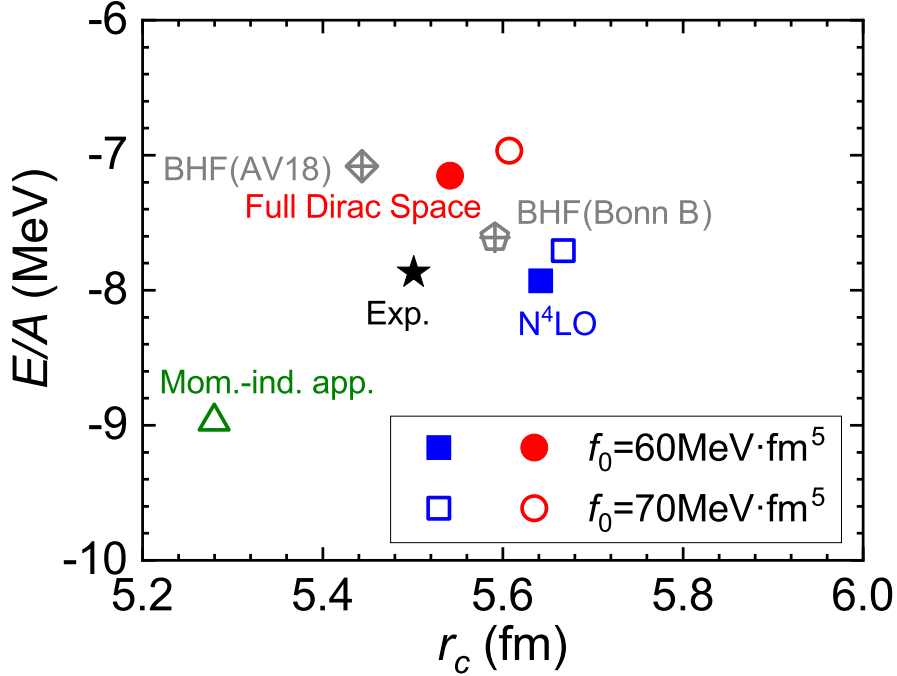}
  \caption{
  (Color online) Energy per nucleon $E/A$ for $^{208}$Pb and the charge radius $r_c$ calculated by the RBHF theory in the full Dirac space, in comparison with the results obtained by the BHF theory~\cite{Bu2016CPL} with the Argonne V18 or Bonn B two-nucleon potentials plus their corresponding microscopic three-nucleon forces, the RBHF theory within the momentum-independence approximation~\cite{Alonso2003PRC}, and the EFT including only two-neutron forces at N$^4$LO~\cite{Sammarruca2016PRC}.
  The solid black star represents the experimental data.
  }
  \label{Fig2}
\end{figure}

In Fig.~\ref{Fig1}, the densities of point protons and neutrons are obtained without considering that protons and neutrons are composite particles with extended size.
In order to compare with the experimental signature of the density in $^{208}$Pb by elastic electron scattering, the charge density is calculated by convolution of the proton density with a Gaussian form factor~\cite{Negele1970PRC}
\begin{equation}
    \rho_{\rm ch}(r) = \frac{1}{a\sqrt{\pi}}\displaystyle \int dr' r' \rho_{p}(r)\left[e^{-(r-r')^2/a^2}/r-e^{-(r+r')^2/a^2}/r \right],
\end{equation}
where $a = \sqrt{2/3} \langle r^2\rangle_p^{1/2}$ is the proton size  for spherically symmetric density distributions~\cite{Yao2012PRC}.
Furthermore, the charge radii $r_c$ can be calculated from the charge density $\rho_{\rm ch}(r)$.

Figure~\ref{Fig2} shows the energy per nucleon $E/A$ for $^{208}$Pb and the charge radius $r_c$ calculated by the RBHF theory in the full Dirac space, in comparison with the BHF theory~\cite{Bu2016CPL}, the RBHF theory within the momentum-independence approximation~\cite{Alonso2003PRC}, and the EFT~\cite{Sammarruca2016PRC}.
The solid black star indicates the experimental data.
As for the RBHF theory in the full Dirac space, the prediction from $f_0$ = 60 MeV$\cdot$fm$^5$ is much closer to the experimental data than those from $f_0$ = 70 MeV$\cdot$fm$^5$, since the magnitude of surface term in Eq.~\eqref{equldm} is proportional to $f_0$ and the latter one introduces more repulsion in the present model.
Similar results also apply to chiral EFT.
To specify the importance of the RBHF calculations in the full Dirac space, the prediction from the RBHF theory within the momentum-independence approximation with $f_0$ = 70 MeV$\cdot$fm$^5$ is also shown~\cite{Alonso2003PRC}.
It can be seen that the full Dirac space gives a larger radius than the momentum-independence approximation, while the magnitude of the energy is smaller.
In order to further improve the accuracy of the results from the RBHF theory in the full Dirac space, the mapping from nuclear matter to finite nuclei in a more microscopic way will be treated carefully in the near future, such as LDA~\cite{Negele1970PRC}.
One can note that the energy per nucleon $E/A$ from chiral EFT  agree well with the experimental data~\cite{Sammarruca2016PRC}, since the EOS for the symmetric nuclear matter was obtained from empirically
determined values of characteristic constants in homogeneous matter at saturation and subsaturation~\cite{Alam2014PRC}.
In addition, the predictions from the RBHF theory in the full Dirac space with only two-nucleon potential are compatible with those from the BHF theory with the Argonne V18 (AV18)~\cite{Wiringa-1995-PhysRevC.51.38} and Bonn B~\cite{Machleidt1989_ANP19-189} two-nucleon potentials plus their corresponding microscopic three-nucleon forces~\cite{LI-ZH2008_PRC77-034316,LI-ZH2008_PRC78-028801}. However, the parameter $f_0$ in the BHF theory is optimized by fitting the binding energies of $^{16}$O, $^{40}$Ca, $^{48}$Ca, $^{90}$Zr, $^{114}$Sn, and $^{208}$Pb and the values $f_0 =$\ 35 (50) MeV$\cdot$fm$^5$ for the AV18 (Bonn B) potential are substantially smaller than 60--70 MeV$\cdot$fm$^5$~\cite{Oyamatsu2010PRC}.
It is obvious that the energy will be far away from the experimental data if $f_0 =$\ 60--70 MeV$\cdot$fm$^5$~\cite{Oyamatsu2010PRC} were used in the BHF theory.

\begin{figure}[htbp]
  \centering
  \includegraphics[width=10cm]{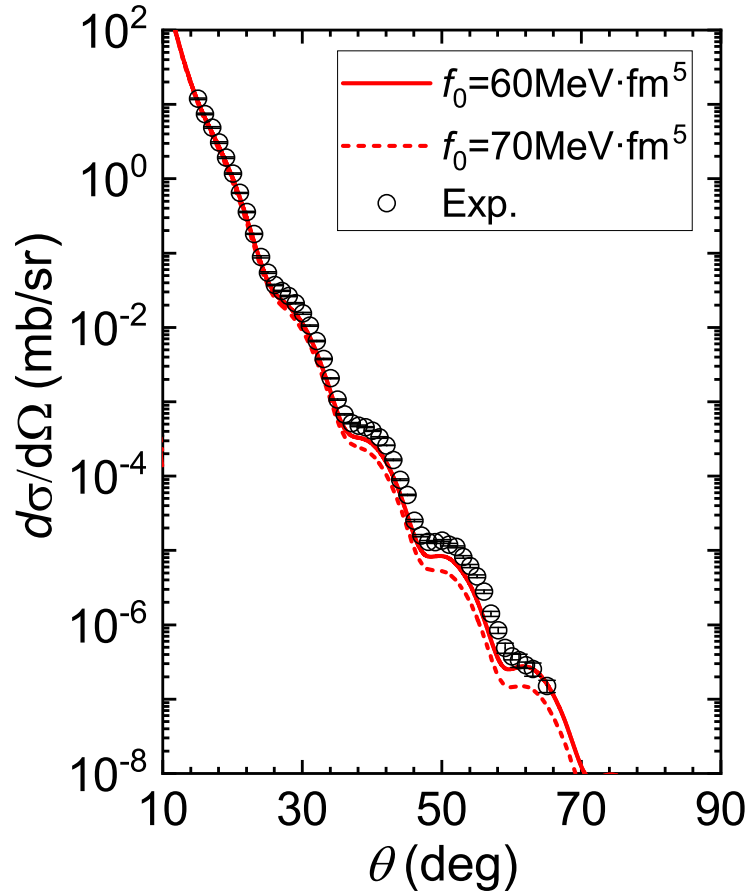}
  \caption{
  (Color online) Elastic DCS for electron-nucleus scattering in $^{208}$Pb as a function of the scattering angle $\theta$ at the electron beam energies 502~MeV.
  The results from the RBHF theory in the full Dirac space are compared with the measured DCS (Exp)~\cite{Friar1973NPA}.
  }
  \label{Fig3}
\end{figure}

The comparison between the charge densities from the RBHF theory in the full Dirac space and the experimental charge densities shall be connected with the discussion of the DCS and the electric charge form factors in the elastic electron-nucleus scattering, which are the quantities measured in real experiments. Several theoretical methods have been used to study the DCS and the electric charge form factors, such as the plane-wave Born approximation, the eikonal approximation, and the phase-shift analysis method~\cite{Baker1964PR,Yennie1954PR}.
In this work, the DCS are calculated by the phase-shift analysis method which has been widely used in elastic electron-nucleus scattering off both stable and unstable
nuclei~\cite{Antonov2005PRC,Sarriguren2007PRC,Roca-Maza2008PRC,Roca-Maza2013PRC}. This method takes into account the Coulomb distortion effect and agrees well with the experimental scattering data in a broad scattering energy range for both light and heavy nuclei~\cite{Chu2009PRC}.

Figure~\ref{Fig3} shows the DCS for elastic electron scattering in $^{208}$Pb as a function of the scattering angle $\theta$ at the electron beam energies 502~MeV.
For $f_0$ = 60 MeV$\cdot$fm$^5$, the results from the RBHF theory give a good description of the experimental data~\cite{Friar1973NPA} at all scattering angles. However, in the case of 70 MeV$\cdot$fm$^5$, the experimental data can only be similarly reproduced at small scattering angles, up to the second diffraction minimum.
The DCS curve obtained from the RBHF theory using $f_0$ = 70 MeV$\cdot$fm$^5$ has the same shape as that from $f_0$ = 60 MeV$\cdot$fm$^5$.
The positions of the diffraction minima and maxima calculated from these two curves are almost the same, but the DCS curve from $f_0$ = 70 MeV$\cdot$fm$^5$ is shifted downward as a whole, especially at large scattering angles.
The deviations for the DCS predicted by the RBHF theory between the two values of $f_0$, and the discrepancies with respect to experimental data, become more prominent with the increasing of the scattering angles.

\begin{figure}[htbp]
  \centering
  \includegraphics[width=10.0cm]{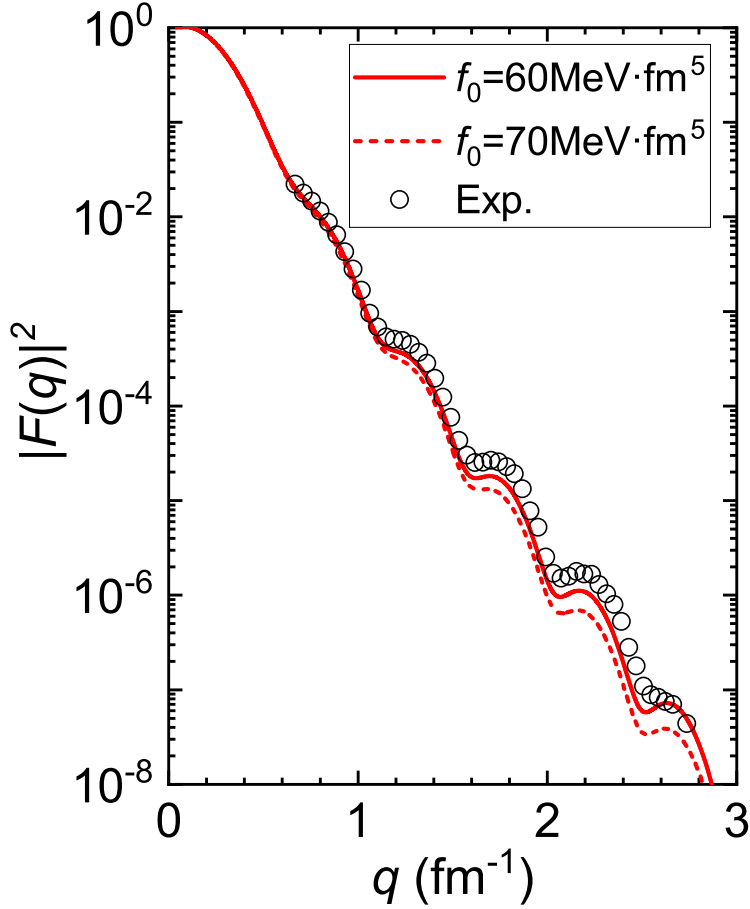}
  \caption{
  (Color online) Squared charge form factor for $^{208}$Pb as a function of the momentum transfer $q$ at the electron beam energies 502~MeV.
  }
  \label{Fig4}
\end{figure}

The electric charge form factor $F(q)$ is a very important quantity to characterize the elastic electron-nucleus scattering which can be used to analyse the effect of the finite size of the nucleus.
At a given beam energy, the squared charge form factor $|F(q)|^2$ is obtained as
\begin{equation}\label{eq:FormFactor}
  |F(q)|^2=\frac{{d\sigma}/{d\Omega}}{{d\sigma_M}/{d\Omega}},
\end{equation}
where the momentum transfer $q$ is related to the scattering angle $\theta$ in the laboratory frame by $q= 2E\sin(\theta/2)$, ${d\sigma}/{d\Omega}$ is the DCS calculated from the RBHF theory in the full Dirac space and ${d\sigma_M}/{d\Omega}$ is the Mott DCS.

In Fig.~\ref{Fig4}, the squared charge form factor $|F(q)|^2$ for the nuclei $^{208}$Pb at the electron beam energies 502~MeV are depicted as functions of the momentum transfer $q$ calculated by the RBHF theory in the full Dirac space.
The experimental data of $|F(q)|^2$ can be well reproduced in the low-momentum transfer regions by the RBHF theory for both $f_0$ = 60 MeV$\cdot$fm$^5$ and $f_0$ = 70 MeV$\cdot$fm$^5$, while the discrepancies appear between the theoretical predictions from the RBHF theory and the experimental data at large-momentum transfer regions, as is expected from the previous analysis of the DCS at small scattering angles in Fig.~\ref{Fig3}.
This indicates that the RBHF theory describe differently the central region of the experimental charge density.
Comparing to $|F(q)|^2$ obtained from the RBHF theory by using $f_0$ = 70 MeV$\cdot$fm$^5$, the positions of the diffraction minima and maxima calculated from $f_0$ = 60 MeV$\cdot$fm$^5$ are almost the same, but the results from $f_0$ = 60 MeV$\cdot$fm$^5$ shifted upward as a whole and became much closer to the experimental data, especially at large-momentum transfer regions.

\section{Summary}\label{SecIV}

The computation of a heavy nucleus, e.g, $^{208}$Pb, has been out of reach for relativistic \emph{ab initio} theory.
Relativistic Brueckner-Hartree-Fock (RBHF) theory provides a promising approach to achieve this goal.
Comparing with the momentum-independence approximation method and the projection method in the RBHF theory, one of the most significant advantage for the RBHF theory in the full Dirac space is that the momentum dependence of scalar and vector components of the single-particle potentials can be determine uniquely.
In order to apply this new method to study finite nuclei, as a first step, properties of $^{208}$Pb are explored by using the microscopic equation of state for asymmetric nuclear matter and a liquid droplet model.
The neutron and proton density distributions, the binding energy and the charge radius are calculated by minimization of the energy.
As for $f_0$ = 60~(70) MeV$\cdot$fm$^5$ in the surface term of the liquid droplet model, the neutron radii, the proton radii and the neutron skin thickness are 5.59~(5.67)~fm, 5.48~(5.55)~fm and 0.11~(0.12)~fm, respectively.
To further compare the charge densities predicted from the RBHF theory in the full Dirac space with the experimental charge densities, the differential cross sections and the electric charge form factors in the elastic electron-nucleus scattering are obtained by using the phase-shift analysis method.
The results from the RBHF theory are in good agreement with the experimental data.
This work will further motivate us to study finite nuclei in a more microscopic way based on the RBHF theory in the full Dirac space in the near future, such as the local density approximation (LDA).

\begin{acknowledgments}

We thank Dr.~Qiang Zhao and Di Wu for helpful discussions.
This work was partly supported by the National Natural Science Foundation of China (NSFC) under Grants No. 12205030 and No. 12147102, the Fundamental Research Funds for the Central Universities under Grants No. 2020CDJQY-Z003 and No. 2021CDJZYJH-003, the MOST-RIKEN Joint Project “\emph{Ab initio} investigation in nuclear physics”.
Part of this work was achieved by using the supercomputer OCTOPUS at the Cybermedia Center, Osaka University under the support of Research Center for Nuclear Physics of Osaka University.

\end{acknowledgments}

\bibliography{RBHF}

\begin{thebibliography}{63}%
\makeatletter
\providecommand \@ifxundefined [1]{%
 \@ifx{#1\undefined}
}%
\providecommand \@ifnum [1]{%
 \ifnum #1\expandafter \@firstoftwo
 \else \expandafter \@secondoftwo
 \fi
}%
\providecommand \@ifx [1]{%
 \ifx #1\expandafter \@firstoftwo
 \else \expandafter \@secondoftwo
 \fi
}%
\providecommand \natexlab [1]{#1}%
\providecommand \enquote  [1]{``#1''}%
\providecommand \bibnamefont  [1]{#1}%
\providecommand \bibfnamefont [1]{#1}%
\providecommand \citenamefont [1]{#1}%
\providecommand \href@noop [0]{\@secondoftwo}%
\providecommand \href [0]{\begingroup \@sanitize@url \@href}%
\providecommand \@href[1]{\@@startlink{#1}\@@href}%
\providecommand \@@href[1]{\endgroup#1\@@endlink}%
\providecommand \@sanitize@url [0]{\catcode `\\12\catcode `\$12\catcode
  `\&12\catcode `\#12\catcode `\^12\catcode `\_12\catcode `\%12\relax}%
\providecommand \@@startlink[1]{}%
\providecommand \@@endlink[0]{}%
\providecommand \url  [0]{\begingroup\@sanitize@url \@url }%
\providecommand \@url [1]{\endgroup\@href {#1}{\urlprefix }}%
\providecommand \urlprefix  [0]{URL }%
\providecommand \Eprint [0]{\href }%
\providecommand \doibase [0]{http://dx.doi.org/}%
\providecommand \selectlanguage [0]{\@gobble}%
\providecommand \bibinfo  [0]{\@secondoftwo}%
\providecommand \bibfield  [0]{\@secondoftwo}%
\providecommand \translation [1]{[#1]}%
\providecommand \BibitemOpen [0]{}%
\providecommand \bibitemStop [0]{}%
\providecommand \bibitemNoStop [0]{.\EOS\space}%
\providecommand \EOS [0]{\spacefactor3000\relax}%
\providecommand \BibitemShut  [1]{\csname bibitem#1\endcsname}%
\let\auto@bib@innerbib\@empty
\bibitem [{\citenamefont {Lattimer}\ and\ \citenamefont
  {Prakash}(2000)}]{Lattimer_2000_PR333_121}%
  \BibitemOpen
  \bibfield  {author} {\bibinfo {author} {\bibfnamefont {J.~M.}\ \bibnamefont
  {Lattimer}}\ and\ \bibinfo {author} {\bibfnamefont {M.}~\bibnamefont
  {Prakash}},\ }\href
  {http://www.sciencedirect.com/science/article/pii/S0370157300000193}
  {\bibfield  {journal} {\bibinfo  {journal} {Phys. Rep.}\ }\textbf {\bibinfo
  {volume} {333}},\ \bibinfo {pages} {121 } (\bibinfo {year}
  {2000})}\BibitemShut {NoStop}%
\bibitem [{\citenamefont {Li}\ \emph {et~al.}(2008{\natexlab{a}})\citenamefont
  {Li}, \citenamefont {Chen},\ and\ \citenamefont {Ko}}]{Li_2008_PR464_113}%
  \BibitemOpen
  \bibfield  {author} {\bibinfo {author} {\bibfnamefont {B.-A.}\ \bibnamefont
  {Li}}, \bibinfo {author} {\bibfnamefont {L.-W.}\ \bibnamefont {Chen}}, \ and\
  \bibinfo {author} {\bibfnamefont {C.~M.}\ \bibnamefont {Ko}},\ }\href
  {http://www.sciencedirect.com/science/article/pii/S0370157308001269}
  {\bibfield  {journal} {\bibinfo  {journal} {Phys. Rep.}\ }\textbf {\bibinfo
  {volume} {464}},\ \bibinfo {pages} {113 } (\bibinfo {year}
  {2008}{\natexlab{a}})}\BibitemShut {NoStop}%
\bibitem [{\citenamefont {Oertel}\ \emph {et~al.}(2017)\citenamefont {Oertel},
  \citenamefont {Hempel}, \citenamefont {Kl\"ahn},\ and\ \citenamefont
  {Typel}}]{Oertel2017RMP_89_015007}%
  \BibitemOpen
  \bibfield  {author} {\bibinfo {author} {\bibfnamefont {M.}~\bibnamefont
  {Oertel}}, \bibinfo {author} {\bibfnamefont {M.}~\bibnamefont {Hempel}},
  \bibinfo {author} {\bibfnamefont {T.}~\bibnamefont {Kl\"ahn}}, \ and\
  \bibinfo {author} {\bibfnamefont {S.}~\bibnamefont {Typel}},\ }\href
  {\doibase 10.1103/RevModPhys.89.015007} {\bibfield  {journal} {\bibinfo
  {journal} {Review of Modern Physics}\ }\textbf {\bibinfo {volume} {89}},\
  \bibinfo {pages} {015007} (\bibinfo {year} {2017})}\BibitemShut {NoStop}%
\bibitem [{\citenamefont {Burgio}\ \emph {et~al.}(2021)\citenamefont {Burgio},
  \citenamefont {Schulze}, \citenamefont {Vidaña},\ and\ \citenamefont
  {Wei}}]{Burgio2021PPNP_120_103879}%
  \BibitemOpen
  \bibfield  {author} {\bibinfo {author} {\bibfnamefont {G.}~\bibnamefont
  {Burgio}}, \bibinfo {author} {\bibfnamefont {H.-J.}\ \bibnamefont {Schulze}},
  \bibinfo {author} {\bibfnamefont {I.}~\bibnamefont {Vidaña}}, \ and\
  \bibinfo {author} {\bibfnamefont {J.-B.}\ \bibnamefont {Wei}},\ }\href
  {\doibase https://doi.org/10.1016/j.ppnp.2021.103879} {\bibfield  {journal}
  {\bibinfo  {journal} {Progress in Particle and Nuclear Physics}\ }\textbf
  {\bibinfo {volume} {120}},\ \bibinfo {pages} {103879} (\bibinfo {year}
  {2021})}\BibitemShut {NoStop}%
\bibitem [{\citenamefont {Alex~Brown}(2000)}]{ABBrown_2000PRL-85-5296}%
  \BibitemOpen
  \bibfield  {author} {\bibinfo {author} {\bibfnamefont {B.}~\bibnamefont
  {Alex~Brown}},\ }\href {\doibase 10.1103/PhysRevLett.85.5296} {\bibfield
  {journal} {\bibinfo  {journal} {Phys. Rev. Lett.}\ }\textbf {\bibinfo
  {volume} {85}},\ \bibinfo {pages} {5296} (\bibinfo {year}
  {2000})}\BibitemShut {NoStop}%
\bibitem [{\citenamefont {Centelles}\ \emph {et~al.}(2009)\citenamefont
  {Centelles}, \citenamefont {Roca-Maza}, \citenamefont {Vi\~nas},\ and\
  \citenamefont {Warda}}]{Centelles_2009_PRL102_122502}%
  \BibitemOpen
  \bibfield  {author} {\bibinfo {author} {\bibfnamefont {M.}~\bibnamefont
  {Centelles}}, \bibinfo {author} {\bibfnamefont {X.}~\bibnamefont
  {Roca-Maza}}, \bibinfo {author} {\bibfnamefont {X.}~\bibnamefont {Vi\~nas}},
  \ and\ \bibinfo {author} {\bibfnamefont {M.}~\bibnamefont {Warda}},\ }\href
  {\doibase 10.1103/PhysRevLett.102.122502} {\bibfield  {journal} {\bibinfo
  {journal} {Phys. Rev. Lett.}\ }\textbf {\bibinfo {volume} {102}},\ \bibinfo
  {pages} {122502} (\bibinfo {year} {2009})}\BibitemShut {NoStop}%
\bibitem [{\citenamefont {Roca-Maza}\ \emph
  {et~al.}(2011{\natexlab{a}})\citenamefont {Roca-Maza}, \citenamefont
  {Centelles}, \citenamefont {Vi\~nas},\ and\ \citenamefont
  {Warda}}]{Roca-Maza_2011-PRL-106-252501}%
  \BibitemOpen
  \bibfield  {author} {\bibinfo {author} {\bibfnamefont {X.}~\bibnamefont
  {Roca-Maza}}, \bibinfo {author} {\bibfnamefont {M.}~\bibnamefont
  {Centelles}}, \bibinfo {author} {\bibfnamefont {X.}~\bibnamefont {Vi\~nas}},
  \ and\ \bibinfo {author} {\bibfnamefont {M.}~\bibnamefont {Warda}},\ }\href
  {\doibase 10.1103/PhysRevLett.106.252501} {\bibfield  {journal} {\bibinfo
  {journal} {Phys. Rev. Lett.}\ }\textbf {\bibinfo {volume} {106}},\ \bibinfo
  {pages} {252501} (\bibinfo {year} {2011}{\natexlab{a}})}\BibitemShut
  {NoStop}%
\bibitem [{\citenamefont {Reed}\ \emph {et~al.}(2021)\citenamefont {Reed},
  \citenamefont {Fattoyev}, \citenamefont {Horowitz},\ and\ \citenamefont
  {Piekarewicz}}]{Reed2021PRL}%
  \BibitemOpen
  \bibfield  {author} {\bibinfo {author} {\bibfnamefont {B.~T.}\ \bibnamefont
  {Reed}}, \bibinfo {author} {\bibfnamefont {F.~J.}\ \bibnamefont {Fattoyev}},
  \bibinfo {author} {\bibfnamefont {C.~J.}\ \bibnamefont {Horowitz}}, \ and\
  \bibinfo {author} {\bibfnamefont {J.}~\bibnamefont {Piekarewicz}},\ }\href
  {\doibase 10.1103/PhysRevLett.126.172503} {\bibfield  {journal} {\bibinfo
  {journal} {Phys. Rev. Lett.}\ }\textbf {\bibinfo {volume} {126}},\ \bibinfo
  {pages} {172503} (\bibinfo {year} {2021})}\BibitemShut {NoStop}%
\bibitem [{\citenamefont {Thiel}\ \emph {et~al.}(2019)\citenamefont {Thiel},
  \citenamefont {Sfienti}, \citenamefont {Piekarewicz}, \citenamefont
  {Horowitz},\ and\ \citenamefont {Vanderhaeghen}}]{Thiel2019JPG}%
  \BibitemOpen
  \bibfield  {author} {\bibinfo {author} {\bibfnamefont {M.}~\bibnamefont
  {Thiel}}, \bibinfo {author} {\bibfnamefont {C.}~\bibnamefont {Sfienti}},
  \bibinfo {author} {\bibfnamefont {J.}~\bibnamefont {Piekarewicz}}, \bibinfo
  {author} {\bibfnamefont {C.~J.}\ \bibnamefont {Horowitz}}, \ and\ \bibinfo
  {author} {\bibfnamefont {M.}~\bibnamefont {Vanderhaeghen}},\ }\href {\doibase
  10.1088/1361-6471/ab2c6d} {\bibfield  {journal} {\bibinfo  {journal} {Journal
  of Physics G: Nuclear and Particle Physics}\ }\textbf {\bibinfo {volume}
  {46}},\ \bibinfo {pages} {093003} (\bibinfo {year} {2019})}\BibitemShut
  {NoStop}%
\bibitem [{\citenamefont {Donnelly}\ \emph {et~al.}(1989)\citenamefont
  {Donnelly}, \citenamefont {Dubach},\ and\ \citenamefont
  {Sick}}]{Donnelly1989NPA}%
  \BibitemOpen
  \bibfield  {author} {\bibinfo {author} {\bibfnamefont {T.}~\bibnamefont
  {Donnelly}}, \bibinfo {author} {\bibfnamefont {J.}~\bibnamefont {Dubach}}, \
  and\ \bibinfo {author} {\bibfnamefont {I.}~\bibnamefont {Sick}},\ }\href
  {\doibase https://doi.org/10.1016/0375-9474(89)90432-6} {\bibfield  {journal}
  {\bibinfo  {journal} {Nuclear Physics A}\ }\textbf {\bibinfo {volume}
  {503}},\ \bibinfo {pages} {589} (\bibinfo {year} {1989})}\BibitemShut
  {NoStop}%
\bibitem [{\citenamefont {Horowitz}\ \emph {et~al.}(2001)\citenamefont
  {Horowitz}, \citenamefont {Pollock}, \citenamefont {Souder},\ and\
  \citenamefont {Michaels}}]{Horowitz2001PRC}%
  \BibitemOpen
  \bibfield  {author} {\bibinfo {author} {\bibfnamefont {C.~J.}\ \bibnamefont
  {Horowitz}}, \bibinfo {author} {\bibfnamefont {S.~J.}\ \bibnamefont
  {Pollock}}, \bibinfo {author} {\bibfnamefont {P.~A.}\ \bibnamefont {Souder}},
  \ and\ \bibinfo {author} {\bibfnamefont {R.}~\bibnamefont {Michaels}},\
  }\href {\doibase 10.1103/PhysRevC.63.025501} {\bibfield  {journal} {\bibinfo
  {journal} {Phys. Rev. C}\ }\textbf {\bibinfo {volume} {63}},\ \bibinfo
  {pages} {025501} (\bibinfo {year} {2001})}\BibitemShut {NoStop}%
\bibitem [{\citenamefont {Adhikari}\ \emph {et~al.}(2021)\citenamefont
  {Adhikari}, \citenamefont {Albataineh}, \citenamefont {Androic},
  \citenamefont {Aniol}, \citenamefont {Armstrong}, \citenamefont {Averett},
  \citenamefont {Ayerbe~Gayoso}, \citenamefont {Barcus}, \citenamefont
  {Bellini}, \citenamefont {Beminiwattha}, \citenamefont {Benesch},
  \citenamefont {Bhatt}, \citenamefont {Bhatta~Pathak}, \citenamefont
  {Bhetuwal}, \citenamefont {Blaikie}, \citenamefont {Campagna}, \citenamefont
  {Camsonne}, \citenamefont {Cates}, \citenamefont {Chen}, \citenamefont
  {Clarke}, \citenamefont {Cornejo}, \citenamefont {Covrig~Dusa}, \citenamefont
  {Datta}, \citenamefont {Deshpande}, \citenamefont {Dutta}, \citenamefont
  {Feldman}, \citenamefont {Fuchey}, \citenamefont {Gal}, \citenamefont
  {Gaskell}, \citenamefont {Gautam}, \citenamefont {Gericke}, \citenamefont
  {Ghosh}, \citenamefont {Halilovic}, \citenamefont {Hansen}, \citenamefont
  {Hauenstein}, \citenamefont {Henry}, \citenamefont {Horowitz}, \citenamefont
  {Jantzi}, \citenamefont {Jian}, \citenamefont {Johnston}, \citenamefont
  {Jones}, \citenamefont {Karki}, \citenamefont {Katugampola}, \citenamefont
  {Keppel}, \citenamefont {King}, \citenamefont {King}, \citenamefont {Knauss},
  \citenamefont {Kumar}, \citenamefont {Kutz}, \citenamefont
  {Lashley-Colthirst}, \citenamefont {Leverick}, \citenamefont {Liu},
  \citenamefont {Liyange}, \citenamefont {Malace}, \citenamefont {Mammei},
  \citenamefont {Mammei}, \citenamefont {McCaughan}, \citenamefont {McNulty},
  \citenamefont {Meekins}, \citenamefont {Metts}, \citenamefont {Michaels},
  \citenamefont {Mondal}, \citenamefont {Napolitano}, \citenamefont {Narayan},
  \citenamefont {Nikolaev}, \citenamefont {Rashad}, \citenamefont {Owen},
  \citenamefont {Palatchi}, \citenamefont {Pan}, \citenamefont {Pandey},
  \citenamefont {Park}, \citenamefont {Paschke}, \citenamefont {Petrusky},
  \citenamefont {Pitt}, \citenamefont {Premathilake}, \citenamefont {Puckett},
  \citenamefont {Quinn}, \citenamefont {Radloff}, \citenamefont {Rahman},
  \citenamefont {Rathnayake}, \citenamefont {Reed}, \citenamefont {Reimer},
  \citenamefont {Richards}, \citenamefont {Riordan}, \citenamefont {Roblin},
  \citenamefont {Seeds}, \citenamefont {Shahinyan}, \citenamefont {Souder},
  \citenamefont {Tang}, \citenamefont {Thiel}, \citenamefont {Tian},
  \citenamefont {Urciuoli}, \citenamefont {Wertz}, \citenamefont
  {Wojtsekhowski}, \citenamefont {Yale}, \citenamefont {Ye}, \citenamefont
  {Yoon}, \citenamefont {Zec}, \citenamefont {Zhang}, \citenamefont {Zhang},\
  and\ \citenamefont {Zheng}}]{Adhikari2021PRL}%
  \BibitemOpen
  \bibfield  {author} {\bibinfo {author} {\bibfnamefont {D.}~\bibnamefont
  {Adhikari}}, \bibinfo {author} {\bibfnamefont {H.}~\bibnamefont
  {Albataineh}}, \bibinfo {author} {\bibfnamefont {D.}~\bibnamefont {Androic}},
  \bibinfo {author} {\bibfnamefont {K.}~\bibnamefont {Aniol}}, \bibinfo
  {author} {\bibfnamefont {D.~S.}\ \bibnamefont {Armstrong}}, \bibinfo {author}
  {\bibfnamefont {T.}~\bibnamefont {Averett}}, \bibinfo {author} {\bibfnamefont
  {C.}~\bibnamefont {Ayerbe~Gayoso}}, \bibinfo {author} {\bibfnamefont
  {S.}~\bibnamefont {Barcus}}, \bibinfo {author} {\bibfnamefont
  {V.}~\bibnamefont {Bellini}}, \bibinfo {author} {\bibfnamefont {R.~S.}\
  \bibnamefont {Beminiwattha}}, \bibinfo {author} {\bibfnamefont {J.~F.}\
  \bibnamefont {Benesch}}, \bibinfo {author} {\bibfnamefont {H.}~\bibnamefont
  {Bhatt}}, \bibinfo {author} {\bibfnamefont {D.}~\bibnamefont
  {Bhatta~Pathak}}, \bibinfo {author} {\bibfnamefont {D.}~\bibnamefont
  {Bhetuwal}}, \bibinfo {author} {\bibfnamefont {B.}~\bibnamefont {Blaikie}},
  \bibinfo {author} {\bibfnamefont {Q.}~\bibnamefont {Campagna}}, \bibinfo
  {author} {\bibfnamefont {A.}~\bibnamefont {Camsonne}}, \bibinfo {author}
  {\bibfnamefont {G.~D.}\ \bibnamefont {Cates}}, \bibinfo {author}
  {\bibfnamefont {Y.}~\bibnamefont {Chen}}, \bibinfo {author} {\bibfnamefont
  {C.}~\bibnamefont {Clarke}}, \bibinfo {author} {\bibfnamefont {J.~C.}\
  \bibnamefont {Cornejo}}, \bibinfo {author} {\bibfnamefont {S.}~\bibnamefont
  {Covrig~Dusa}}, \bibinfo {author} {\bibfnamefont {P.}~\bibnamefont {Datta}},
  \bibinfo {author} {\bibfnamefont {A.}~\bibnamefont {Deshpande}}, \bibinfo
  {author} {\bibfnamefont {D.}~\bibnamefont {Dutta}}, \bibinfo {author}
  {\bibfnamefont {C.}~\bibnamefont {Feldman}}, \bibinfo {author} {\bibfnamefont
  {E.}~\bibnamefont {Fuchey}}, \bibinfo {author} {\bibfnamefont
  {C.}~\bibnamefont {Gal}}, \bibinfo {author} {\bibfnamefont {D.}~\bibnamefont
  {Gaskell}}, \bibinfo {author} {\bibfnamefont {T.}~\bibnamefont {Gautam}},
  \bibinfo {author} {\bibfnamefont {M.}~\bibnamefont {Gericke}}, \bibinfo
  {author} {\bibfnamefont {C.}~\bibnamefont {Ghosh}}, \bibinfo {author}
  {\bibfnamefont {I.}~\bibnamefont {Halilovic}}, \bibinfo {author}
  {\bibfnamefont {J.-O.}\ \bibnamefont {Hansen}}, \bibinfo {author}
  {\bibfnamefont {F.}~\bibnamefont {Hauenstein}}, \bibinfo {author}
  {\bibfnamefont {W.}~\bibnamefont {Henry}}, \bibinfo {author} {\bibfnamefont
  {C.~J.}\ \bibnamefont {Horowitz}}, \bibinfo {author} {\bibfnamefont
  {C.}~\bibnamefont {Jantzi}}, \bibinfo {author} {\bibfnamefont
  {S.}~\bibnamefont {Jian}}, \bibinfo {author} {\bibfnamefont {S.}~\bibnamefont
  {Johnston}}, \bibinfo {author} {\bibfnamefont {D.~C.}\ \bibnamefont {Jones}},
  \bibinfo {author} {\bibfnamefont {B.}~\bibnamefont {Karki}}, \bibinfo
  {author} {\bibfnamefont {S.}~\bibnamefont {Katugampola}}, \bibinfo {author}
  {\bibfnamefont {C.}~\bibnamefont {Keppel}}, \bibinfo {author} {\bibfnamefont
  {P.~M.}\ \bibnamefont {King}}, \bibinfo {author} {\bibfnamefont {D.~E.}\
  \bibnamefont {King}}, \bibinfo {author} {\bibfnamefont {M.}~\bibnamefont
  {Knauss}}, \bibinfo {author} {\bibfnamefont {K.~S.}\ \bibnamefont {Kumar}},
  \bibinfo {author} {\bibfnamefont {T.}~\bibnamefont {Kutz}}, \bibinfo {author}
  {\bibfnamefont {N.}~\bibnamefont {Lashley-Colthirst}}, \bibinfo {author}
  {\bibfnamefont {G.}~\bibnamefont {Leverick}}, \bibinfo {author}
  {\bibfnamefont {H.}~\bibnamefont {Liu}}, \bibinfo {author} {\bibfnamefont
  {N.}~\bibnamefont {Liyange}}, \bibinfo {author} {\bibfnamefont
  {S.}~\bibnamefont {Malace}}, \bibinfo {author} {\bibfnamefont
  {R.}~\bibnamefont {Mammei}}, \bibinfo {author} {\bibfnamefont
  {J.}~\bibnamefont {Mammei}}, \bibinfo {author} {\bibfnamefont
  {M.}~\bibnamefont {McCaughan}}, \bibinfo {author} {\bibfnamefont
  {D.}~\bibnamefont {McNulty}}, \bibinfo {author} {\bibfnamefont
  {D.}~\bibnamefont {Meekins}}, \bibinfo {author} {\bibfnamefont
  {C.}~\bibnamefont {Metts}}, \bibinfo {author} {\bibfnamefont
  {R.}~\bibnamefont {Michaels}}, \bibinfo {author} {\bibfnamefont {M.~M.}\
  \bibnamefont {Mondal}}, \bibinfo {author} {\bibfnamefont {J.}~\bibnamefont
  {Napolitano}}, \bibinfo {author} {\bibfnamefont {A.}~\bibnamefont {Narayan}},
  \bibinfo {author} {\bibfnamefont {D.}~\bibnamefont {Nikolaev}}, \bibinfo
  {author} {\bibfnamefont {M.~N.~H.}\ \bibnamefont {Rashad}}, \bibinfo {author}
  {\bibfnamefont {V.}~\bibnamefont {Owen}}, \bibinfo {author} {\bibfnamefont
  {C.}~\bibnamefont {Palatchi}}, \bibinfo {author} {\bibfnamefont
  {J.}~\bibnamefont {Pan}}, \bibinfo {author} {\bibfnamefont {B.}~\bibnamefont
  {Pandey}}, \bibinfo {author} {\bibfnamefont {S.}~\bibnamefont {Park}},
  \bibinfo {author} {\bibfnamefont {K.~D.}\ \bibnamefont {Paschke}}, \bibinfo
  {author} {\bibfnamefont {M.}~\bibnamefont {Petrusky}}, \bibinfo {author}
  {\bibfnamefont {M.~L.}\ \bibnamefont {Pitt}}, \bibinfo {author}
  {\bibfnamefont {S.}~\bibnamefont {Premathilake}}, \bibinfo {author}
  {\bibfnamefont {A.~J.~R.}\ \bibnamefont {Puckett}}, \bibinfo {author}
  {\bibfnamefont {B.}~\bibnamefont {Quinn}}, \bibinfo {author} {\bibfnamefont
  {R.}~\bibnamefont {Radloff}}, \bibinfo {author} {\bibfnamefont
  {S.}~\bibnamefont {Rahman}}, \bibinfo {author} {\bibfnamefont
  {A.}~\bibnamefont {Rathnayake}}, \bibinfo {author} {\bibfnamefont {B.~T.}\
  \bibnamefont {Reed}}, \bibinfo {author} {\bibfnamefont {P.~E.}\ \bibnamefont
  {Reimer}}, \bibinfo {author} {\bibfnamefont {R.}~\bibnamefont {Richards}},
  \bibinfo {author} {\bibfnamefont {S.}~\bibnamefont {Riordan}}, \bibinfo
  {author} {\bibfnamefont {Y.}~\bibnamefont {Roblin}}, \bibinfo {author}
  {\bibfnamefont {S.}~\bibnamefont {Seeds}}, \bibinfo {author} {\bibfnamefont
  {A.}~\bibnamefont {Shahinyan}}, \bibinfo {author} {\bibfnamefont
  {P.}~\bibnamefont {Souder}}, \bibinfo {author} {\bibfnamefont
  {L.}~\bibnamefont {Tang}}, \bibinfo {author} {\bibfnamefont {M.}~\bibnamefont
  {Thiel}}, \bibinfo {author} {\bibfnamefont {Y.}~\bibnamefont {Tian}},
  \bibinfo {author} {\bibfnamefont {G.~M.}\ \bibnamefont {Urciuoli}}, \bibinfo
  {author} {\bibfnamefont {E.~W.}\ \bibnamefont {Wertz}}, \bibinfo {author}
  {\bibfnamefont {B.}~\bibnamefont {Wojtsekhowski}}, \bibinfo {author}
  {\bibfnamefont {B.}~\bibnamefont {Yale}}, \bibinfo {author} {\bibfnamefont
  {T.}~\bibnamefont {Ye}}, \bibinfo {author} {\bibfnamefont {A.}~\bibnamefont
  {Yoon}}, \bibinfo {author} {\bibfnamefont {A.}~\bibnamefont {Zec}}, \bibinfo
  {author} {\bibfnamefont {W.}~\bibnamefont {Zhang}}, \bibinfo {author}
  {\bibfnamefont {J.}~\bibnamefont {Zhang}}, \ and\ \bibinfo {author}
  {\bibfnamefont {X.}~\bibnamefont {Zheng}} (\bibinfo {collaboration} {PREX
  Collaboration}),\ }\href {\doibase 10.1103/PhysRevLett.126.172502} {\bibfield
   {journal} {\bibinfo  {journal} {Phys. Rev. Lett.}\ }\textbf {\bibinfo
  {volume} {126}},\ \bibinfo {pages} {172502} (\bibinfo {year}
  {2021})}\BibitemShut {NoStop}%
\bibitem [{\citenamefont {Roca-Maza}\ \emph
  {et~al.}(2011{\natexlab{b}})\citenamefont {Roca-Maza}, \citenamefont
  {Centelles}, \citenamefont {Vi\~nas},\ and\ \citenamefont
  {Warda}}]{Roca-Maza2011_PRL-106-252501}%
  \BibitemOpen
  \bibfield  {author} {\bibinfo {author} {\bibfnamefont {X.}~\bibnamefont
  {Roca-Maza}}, \bibinfo {author} {\bibfnamefont {M.}~\bibnamefont
  {Centelles}}, \bibinfo {author} {\bibfnamefont {X.}~\bibnamefont {Vi\~nas}},
  \ and\ \bibinfo {author} {\bibfnamefont {M.}~\bibnamefont {Warda}},\ }\href
  {\doibase 10.1103/PhysRevLett.106.252501} {\bibfield  {journal} {\bibinfo
  {journal} {Phys. Rev. Lett.}\ }\textbf {\bibinfo {volume} {106}},\ \bibinfo
  {pages} {252501} (\bibinfo {year} {2011}{\natexlab{b}})}\BibitemShut
  {NoStop}%
\bibitem [{\citenamefont {Hu}\ \emph {et~al.}(2022)\citenamefont {Hu},
  \citenamefont {Jiang}, \citenamefont {Miyagi}, \citenamefont {Sun},
  \citenamefont {Ekstr\"om}, \citenamefont {Forss\'en}, \citenamefont {Hagen},
  \citenamefont {Holt}, \citenamefont {Papenbrock}, \citenamefont {Stroberg},\
  and\ \citenamefont {Vernon}}]{Hu2022Nature}%
  \BibitemOpen
  \bibfield  {author} {\bibinfo {author} {\bibfnamefont {B.}~\bibnamefont
  {Hu}}, \bibinfo {author} {\bibfnamefont {W.}~\bibnamefont {Jiang}}, \bibinfo
  {author} {\bibfnamefont {T.}~\bibnamefont {Miyagi}}, \bibinfo {author}
  {\bibfnamefont {Z.}~\bibnamefont {Sun}}, \bibinfo {author} {\bibfnamefont
  {A.}~\bibnamefont {Ekstr\"om}}, \bibinfo {author} {\bibfnamefont
  {C.}~\bibnamefont {Forss\'en}}, \bibinfo {author} {\bibfnamefont
  {G.}~\bibnamefont {Hagen}}, \bibinfo {author} {\bibfnamefont {J.~D.}\
  \bibnamefont {Holt}}, \bibinfo {author} {\bibfnamefont {T.}~\bibnamefont
  {Papenbrock}}, \bibinfo {author} {\bibfnamefont {S.~R.}\ \bibnamefont
  {Stroberg}}, \ and\ \bibinfo {author} {\bibfnamefont {I.}~\bibnamefont
  {Vernon}},\ }\href {\doibase 10.1038/s41567-022-01715-8} {\bibfield
  {journal} {\bibinfo  {journal} {Nature Phys.}\ }\textbf {\bibinfo {volume}
  {18}},\ \bibinfo {pages} {1196} (\bibinfo {year} {2022})}\BibitemShut
  {NoStop}%
\bibitem [{\citenamefont {Brockmann}\ and\ \citenamefont
  {Machleidt}(1990)}]{Brockmann1990_PRC42-1965}%
  \BibitemOpen
  \bibfield  {author} {\bibinfo {author} {\bibfnamefont {R.}~\bibnamefont
  {Brockmann}}\ and\ \bibinfo {author} {\bibfnamefont {R.}~\bibnamefont
  {Machleidt}},\ }\href@noop {} {\bibfield  {journal} {\bibinfo  {journal}
  {Phys. Rev. C}\ }\textbf {\bibinfo {volume} {42}},\ \bibinfo {pages} {1965}
  (\bibinfo {year} {1990})}\BibitemShut {NoStop}%
\bibitem [{\citenamefont {Shen}\ \emph {et~al.}(2017)\citenamefont {Shen},
  \citenamefont {Liang}, \citenamefont {Meng}, \citenamefont {Ring},\ and\
  \citenamefont {Zhang}}]{SHEN-SH2017_PRC96-014316}%
  \BibitemOpen
  \bibfield  {author} {\bibinfo {author} {\bibfnamefont {S.}~\bibnamefont
  {Shen}}, \bibinfo {author} {\bibfnamefont {H.}~\bibnamefont {Liang}},
  \bibinfo {author} {\bibfnamefont {J.}~\bibnamefont {Meng}}, \bibinfo {author}
  {\bibfnamefont {P.}~\bibnamefont {Ring}}, \ and\ \bibinfo {author}
  {\bibfnamefont {S.}~\bibnamefont {Zhang}},\ }\href@noop {} {\bibfield
  {journal} {\bibinfo  {journal} {Phys. Rev. C}\ }\textbf {\bibinfo {volume}
  {96}},\ \bibinfo {pages} {014316} (\bibinfo {year} {2017})}\BibitemShut
  {NoStop}%
\bibitem [{\citenamefont {Shen}\ \emph {et~al.}(2019)\citenamefont {Shen},
  \citenamefont {Liang}, \citenamefont {Long}, \citenamefont {Meng},\ and\
  \citenamefont {Ring}}]{SHEN-SH2019_PPNP109-103713}%
  \BibitemOpen
  \bibfield  {author} {\bibinfo {author} {\bibfnamefont {S.}~\bibnamefont
  {Shen}}, \bibinfo {author} {\bibfnamefont {H.}~\bibnamefont {Liang}},
  \bibinfo {author} {\bibfnamefont {W.}~\bibnamefont {Long}}, \bibinfo {author}
  {\bibfnamefont {J.}~\bibnamefont {Meng}}, \ and\ \bibinfo {author}
  {\bibfnamefont {P.}~\bibnamefont {Ring}},\ }\href@noop {} {\bibfield
  {journal} {\bibinfo  {journal} {Prog. Part. Nucl. Phys.}\ }\textbf {\bibinfo
  {volume} {109}},\ \bibinfo {pages} {103713} (\bibinfo {year}
  {2019})}\BibitemShut {NoStop}%
\bibitem [{\citenamefont {Shen}\ \emph {et~al.}(2018)\citenamefont {Shen},
  \citenamefont {Liang}, \citenamefont {Meng}, \citenamefont {Ring},\ and\
  \citenamefont {Zhang}}]{SHEN-SH2018_PRC97-054312}%
  \BibitemOpen
  \bibfield  {author} {\bibinfo {author} {\bibfnamefont {S.}~\bibnamefont
  {Shen}}, \bibinfo {author} {\bibfnamefont {H.}~\bibnamefont {Liang}},
  \bibinfo {author} {\bibfnamefont {J.}~\bibnamefont {Meng}}, \bibinfo {author}
  {\bibfnamefont {P.}~\bibnamefont {Ring}}, \ and\ \bibinfo {author}
  {\bibfnamefont {S.}~\bibnamefont {Zhang}},\ }\href {\doibase
  10.1103/PhysRevC.97.054312} {\bibfield  {journal} {\bibinfo  {journal} {Phys.
  Rev. C}\ }\textbf {\bibinfo {volume} {97}},\ \bibinfo {pages} {054312}
  (\bibinfo {year} {2018})}\BibitemShut {NoStop}%
\bibitem [{\citenamefont {Negele}(1970)}]{Negele1970PRC}%
  \BibitemOpen
  \bibfield  {author} {\bibinfo {author} {\bibfnamefont {J.~W.}\ \bibnamefont
  {Negele}},\ }\href {\doibase 10.1103/PhysRevC.1.1260} {\bibfield  {journal}
  {\bibinfo  {journal} {Phys. Rev. C}\ }\textbf {\bibinfo {volume} {1}},\
  \bibinfo {pages} {1260} (\bibinfo {year} {1970})}\BibitemShut {NoStop}%
\bibitem [{\citenamefont {Oyamatsu}\ \emph {et~al.}(1998)\citenamefont
  {Oyamatsu}, \citenamefont {Tanihata}, \citenamefont {Sugahara}, \citenamefont
  {Sumiyoshi},\ and\ \citenamefont {Toki}}]{OYAMATSU1998NPA}%
  \BibitemOpen
  \bibfield  {author} {\bibinfo {author} {\bibfnamefont {K.}~\bibnamefont
  {Oyamatsu}}, \bibinfo {author} {\bibfnamefont {I.}~\bibnamefont {Tanihata}},
  \bibinfo {author} {\bibfnamefont {Y.}~\bibnamefont {Sugahara}}, \bibinfo
  {author} {\bibfnamefont {K.}~\bibnamefont {Sumiyoshi}}, \ and\ \bibinfo
  {author} {\bibfnamefont {H.}~\bibnamefont {Toki}},\ }\href {\doibase
  https://doi.org/10.1016/S0375-9474(98)00125-0} {\bibfield  {journal}
  {\bibinfo  {journal} {Nuclear Physics A}\ }\textbf {\bibinfo {volume}
  {634}},\ \bibinfo {pages} {3} (\bibinfo {year} {1998})}\BibitemShut {NoStop}%
\bibitem [{\citenamefont {Alonso}\ and\ \citenamefont
  {Sammarruca}(2003)}]{Alonso2003PRC}%
  \BibitemOpen
  \bibfield  {author} {\bibinfo {author} {\bibfnamefont {D.}~\bibnamefont
  {Alonso}}\ and\ \bibinfo {author} {\bibfnamefont {F.}~\bibnamefont
  {Sammarruca}},\ }\href {\doibase 10.1103/PhysRevC.68.054305} {\bibfield
  {journal} {\bibinfo  {journal} {Phys. Rev. C}\ }\textbf {\bibinfo {volume}
  {68}},\ \bibinfo {pages} {054305} (\bibinfo {year} {2003})}\BibitemShut
  {NoStop}%
\bibitem [{\citenamefont {Sammarruca}\ and\ \citenamefont
  {Liu}(2009)}]{Sammarruca2009_PRC79_057301}%
  \BibitemOpen
  \bibfield  {author} {\bibinfo {author} {\bibfnamefont {F.}~\bibnamefont
  {Sammarruca}}\ and\ \bibinfo {author} {\bibfnamefont {P.}~\bibnamefont
  {Liu}},\ }\href {\doibase 10.1103/PhysRevC.79.057301} {\bibfield  {journal}
  {\bibinfo  {journal} {Phys. Rev. C}\ }\textbf {\bibinfo {volume} {79}},\
  \bibinfo {pages} {057301} (\bibinfo {year} {2009})}\BibitemShut {NoStop}%
\bibitem [{\citenamefont {Sammarruca}(2016)}]{Sammarruca2016PRC}%
  \BibitemOpen
  \bibfield  {author} {\bibinfo {author} {\bibfnamefont {F.}~\bibnamefont
  {Sammarruca}},\ }\href {\doibase 10.1103/PhysRevC.94.054317} {\bibfield
  {journal} {\bibinfo  {journal} {Phys. Rev. C}\ }\textbf {\bibinfo {volume}
  {94}},\ \bibinfo {pages} {054317} (\bibinfo {year} {2016})}\BibitemShut
  {NoStop}%
\bibitem [{\citenamefont {Sammarruca}\ and\ \citenamefont
  {Nosyk}(2016)}]{Sammarruca2016_PRC94_044311}%
  \BibitemOpen
  \bibfield  {author} {\bibinfo {author} {\bibfnamefont {F.}~\bibnamefont
  {Sammarruca}}\ and\ \bibinfo {author} {\bibfnamefont {Y.}~\bibnamefont
  {Nosyk}},\ }\href {\doibase 10.1103/PhysRevC.94.044311} {\bibfield  {journal}
  {\bibinfo  {journal} {Phys. Rev. C}\ }\textbf {\bibinfo {volume} {94}},\
  \bibinfo {pages} {044311} (\bibinfo {year} {2016})}\BibitemShut {NoStop}%
\bibitem [{\citenamefont {Gross-Boelting}\ \emph {et~al.}(1999)\citenamefont
  {Gross-Boelting}, \citenamefont {Fuchs},\ and\ \citenamefont
  {Faessler}}]{Gross-Boelting1999_NPA648-105}%
  \BibitemOpen
  \bibfield  {author} {\bibinfo {author} {\bibfnamefont {T.}~\bibnamefont
  {Gross-Boelting}}, \bibinfo {author} {\bibfnamefont {C.}~\bibnamefont
  {Fuchs}}, \ and\ \bibinfo {author} {\bibfnamefont {A.}~\bibnamefont
  {Faessler}},\ }\href@noop {} {\bibfield  {journal} {\bibinfo  {journal}
  {Nuclear Physics A}\ }\textbf {\bibinfo {volume} {648}},\ \bibinfo {pages}
  {105} (\bibinfo {year} {1999})}\BibitemShut {NoStop}%
\bibitem [{\citenamefont {Schiller}\ and\ \citenamefont
  {M\"uther}(2001)}]{Schiller2001_EPJA11-15}%
  \BibitemOpen
  \bibfield  {author} {\bibinfo {author} {\bibfnamefont {E.}~\bibnamefont
  {Schiller}}\ and\ \bibinfo {author} {\bibfnamefont {H.}~\bibnamefont
  {M\"uther}},\ }\href@noop {} {\bibfield  {journal} {\bibinfo  {journal} {Eur.
  Phys. J. A}\ }\textbf {\bibinfo {volume} {11}},\ \bibinfo {pages} {15}
  (\bibinfo {year} {2001})}\BibitemShut {NoStop}%
\bibitem [{\citenamefont {Ulrych}\ and\ \citenamefont
  {M\"uther}(1997)}]{Ulrych1997_PRC56-1788}%
  \BibitemOpen
  \bibfield  {author} {\bibinfo {author} {\bibfnamefont {S.}~\bibnamefont
  {Ulrych}}\ and\ \bibinfo {author} {\bibfnamefont {H.}~\bibnamefont
  {M\"uther}},\ }\href@noop {} {\bibfield  {journal} {\bibinfo  {journal}
  {Phys. Rev. C}\ }\textbf {\bibinfo {volume} {56}},\ \bibinfo {pages} {1788}
  (\bibinfo {year} {1997})}\BibitemShut {NoStop}%
\bibitem [{\citenamefont {Brockmann}\ and\ \citenamefont
  {Toki}(1992)}]{Brockmann1992_PRL68-3408}%
  \BibitemOpen
  \bibfield  {author} {\bibinfo {author} {\bibfnamefont {R.}~\bibnamefont
  {Brockmann}}\ and\ \bibinfo {author} {\bibfnamefont {H.}~\bibnamefont
  {Toki}},\ }\href {\doibase 10.1103/PhysRevLett.68.3408} {\bibfield  {journal}
  {\bibinfo  {journal} {Phys. Rev. Lett.}\ }\textbf {\bibinfo {volume} {68}},\
  \bibinfo {pages} {3408} (\bibinfo {year} {1992})}\BibitemShut {NoStop}%
\bibitem [{\citenamefont {Fritz}\ \emph {et~al.}(1993)\citenamefont {Fritz},
  \citenamefont {M\"uther},\ and\ \citenamefont
  {Machleidt}}]{Fritz1993_PRL71_46}%
  \BibitemOpen
  \bibfield  {author} {\bibinfo {author} {\bibfnamefont {R.}~\bibnamefont
  {Fritz}}, \bibinfo {author} {\bibfnamefont {H.}~\bibnamefont {M\"uther}}, \
  and\ \bibinfo {author} {\bibfnamefont {R.}~\bibnamefont {Machleidt}},\ }\href
  {\doibase 10.1103/PhysRevLett.71.46} {\bibfield  {journal} {\bibinfo
  {journal} {Phys. Rev. Lett.}\ }\textbf {\bibinfo {volume} {71}},\ \bibinfo
  {pages} {46} (\bibinfo {year} {1993})}\BibitemShut {NoStop}%
\bibitem [{\citenamefont {Fuchs}\ \emph {et~al.}(1995)\citenamefont {Fuchs},
  \citenamefont {Lenske},\ and\ \citenamefont {Wolter}}]{Fuchs1995_PRC52_3043}%
  \BibitemOpen
  \bibfield  {author} {\bibinfo {author} {\bibfnamefont {C.}~\bibnamefont
  {Fuchs}}, \bibinfo {author} {\bibfnamefont {H.}~\bibnamefont {Lenske}}, \
  and\ \bibinfo {author} {\bibfnamefont {H.~H.}\ \bibnamefont {Wolter}},\
  }\href {\doibase 10.1103/PhysRevC.52.3043} {\bibfield  {journal} {\bibinfo
  {journal} {Phys. Rev. C}\ }\textbf {\bibinfo {volume} {52}},\ \bibinfo
  {pages} {3043} (\bibinfo {year} {1995})}\BibitemShut {NoStop}%
\bibitem [{\citenamefont {Shen}\ \emph {et~al.}(1997)\citenamefont {Shen},
  \citenamefont {Sugahara},\ and\ \citenamefont
  {Toki}}]{SHEN-H1997_PRC55-1211}%
  \BibitemOpen
  \bibfield  {author} {\bibinfo {author} {\bibfnamefont {H.}~\bibnamefont
  {Shen}}, \bibinfo {author} {\bibfnamefont {Y.}~\bibnamefont {Sugahara}}, \
  and\ \bibinfo {author} {\bibfnamefont {H.}~\bibnamefont {Toki}},\ }\href
  {\doibase 10.1103/PhysRevC.55.1211} {\bibfield  {journal} {\bibinfo
  {journal} {Phys. Rev. C}\ }\textbf {\bibinfo {volume} {55}},\ \bibinfo
  {pages} {1211} (\bibinfo {year} {1997})}\BibitemShut {NoStop}%
\bibitem [{\citenamefont {Ma}\ and\ \citenamefont
  {Liu}(2002)}]{Ma2002_PRC66_024321}%
  \BibitemOpen
  \bibfield  {author} {\bibinfo {author} {\bibfnamefont {Z.-y.}\ \bibnamefont
  {Ma}}\ and\ \bibinfo {author} {\bibfnamefont {L.}~\bibnamefont {Liu}},\
  }\href {\doibase 10.1103/PhysRevC.66.024321} {\bibfield  {journal} {\bibinfo
  {journal} {Phys. Rev. C}\ }\textbf {\bibinfo {volume} {66}},\ \bibinfo
  {pages} {024321} (\bibinfo {year} {2002})}\BibitemShut {NoStop}%
\bibitem [{\citenamefont {van Dalen}\ and\ \citenamefont
  {M\"uther}(2011)}]{vanDalen2011_PRC84_024320}%
  \BibitemOpen
  \bibfield  {author} {\bibinfo {author} {\bibfnamefont {E.~N.~E.}\
  \bibnamefont {van Dalen}}\ and\ \bibinfo {author} {\bibfnamefont
  {H.}~\bibnamefont {M\"uther}},\ }\href {\doibase 10.1103/PhysRevC.84.024320}
  {\bibfield  {journal} {\bibinfo  {journal} {Phys. Rev. C}\ }\textbf {\bibinfo
  {volume} {84}},\ \bibinfo {pages} {024320} (\bibinfo {year}
  {2011})}\BibitemShut {NoStop}%
\bibitem [{\citenamefont {Wang}\ \emph {et~al.}(2021)\citenamefont {Wang},
  \citenamefont {Zhao}, \citenamefont {Ring},\ and\ \citenamefont
  {Meng}}]{WANG-SB2021_PRC103-054319}%
  \BibitemOpen
  \bibfield  {author} {\bibinfo {author} {\bibfnamefont {S.}~\bibnamefont
  {Wang}}, \bibinfo {author} {\bibfnamefont {Q.}~\bibnamefont {Zhao}}, \bibinfo
  {author} {\bibfnamefont {P.}~\bibnamefont {Ring}}, \ and\ \bibinfo {author}
  {\bibfnamefont {J.}~\bibnamefont {Meng}},\ }\href {\doibase
  10.1103/PhysRevC.103.054319} {\bibfield  {journal} {\bibinfo  {journal}
  {Phys. Rev. C}\ }\textbf {\bibinfo {volume} {103}},\ \bibinfo {pages}
  {054319} (\bibinfo {year} {2021})}\BibitemShut {NoStop}%
\bibitem [{\citenamefont {Wang}\ \emph
  {et~al.}(2022{\natexlab{a}})\citenamefont {Wang}, \citenamefont {Tong},
  \citenamefont {Zhao}, \citenamefont {Wang}, \citenamefont {Ring},\ and\
  \citenamefont {Meng}}]{2022-Wang-SIBO-PhysRevC.106.L021305}%
  \BibitemOpen
  \bibfield  {author} {\bibinfo {author} {\bibfnamefont {S.}~\bibnamefont
  {Wang}}, \bibinfo {author} {\bibfnamefont {H.}~\bibnamefont {Tong}}, \bibinfo
  {author} {\bibfnamefont {Q.}~\bibnamefont {Zhao}}, \bibinfo {author}
  {\bibfnamefont {C.}~\bibnamefont {Wang}}, \bibinfo {author} {\bibfnamefont
  {P.}~\bibnamefont {Ring}}, \ and\ \bibinfo {author} {\bibfnamefont
  {J.}~\bibnamefont {Meng}},\ }\href {\doibase 10.1103/PhysRevC.106.L021305}
  {\bibfield  {journal} {\bibinfo  {journal} {Phys. Rev. C}\ }\textbf {\bibinfo
  {volume} {106}},\ \bibinfo {pages} {L021305} (\bibinfo {year}
  {2022}{\natexlab{a}})}\BibitemShut {NoStop}%
\bibitem [{\citenamefont {Wang}\ \emph
  {et~al.}(2022{\natexlab{b}})\citenamefont {Wang}, \citenamefont {Tong},\ and\
  \citenamefont {Wang}}]{WANG-SB2022_PhysRevC.105.054309}%
  \BibitemOpen
  \bibfield  {author} {\bibinfo {author} {\bibfnamefont {S.}~\bibnamefont
  {Wang}}, \bibinfo {author} {\bibfnamefont {H.}~\bibnamefont {Tong}}, \ and\
  \bibinfo {author} {\bibfnamefont {C.}~\bibnamefont {Wang}},\ }\href {\doibase
  10.1103/PhysRevC.105.054309} {\bibfield  {journal} {\bibinfo  {journal}
  {Phys. Rev. C}\ }\textbf {\bibinfo {volume} {105}},\ \bibinfo {pages}
  {054309} (\bibinfo {year} {2022}{\natexlab{b}})}\BibitemShut {NoStop}%
\bibitem [{\citenamefont {Tong}\ \emph {et~al.}(2022)\citenamefont {Tong},
  \citenamefont {Wang},\ and\ \citenamefont
  {Wang}}]{Tong_2022-AstrophysicsJ930.137}%
  \BibitemOpen
  \bibfield  {author} {\bibinfo {author} {\bibfnamefont {H.}~\bibnamefont
  {Tong}}, \bibinfo {author} {\bibfnamefont {C.}~\bibnamefont {Wang}}, \ and\
  \bibinfo {author} {\bibfnamefont {S.}~\bibnamefont {Wang}},\ }\href {\doibase
  10.3847/1538-4357/ac65fc} {\bibfield  {journal} {\bibinfo  {journal} {The
  Astrophysical Journal}\ }\textbf {\bibinfo {volume} {930}},\ \bibinfo {pages}
  {137} (\bibinfo {year} {2022})}\BibitemShut {NoStop}%
\bibitem [{\citenamefont {Wang}\ \emph
  {et~al.}(2022{\natexlab{c}})\citenamefont {Wang}, \citenamefont {Wang},\ and\
  \citenamefont {Tong}}]{Wang2022_PRC106_045804}%
  \BibitemOpen
  \bibfield  {author} {\bibinfo {author} {\bibfnamefont {S.}~\bibnamefont
  {Wang}}, \bibinfo {author} {\bibfnamefont {C.}~\bibnamefont {Wang}}, \ and\
  \bibinfo {author} {\bibfnamefont {H.}~\bibnamefont {Tong}},\ }\href {\doibase
  10.1103/PhysRevC.106.045804} {\bibfield  {journal} {\bibinfo  {journal}
  {Phys. Rev. C}\ }\textbf {\bibinfo {volume} {106}},\ \bibinfo {pages}
  {045804} (\bibinfo {year} {2022}{\natexlab{c}})}\BibitemShut {NoStop}%
\bibitem [{\citenamefont {Serot}\ and\ \citenamefont
  {Walecka}(1986)}]{Serot1986_ANP16-1}%
  \BibitemOpen
  \bibfield  {author} {\bibinfo {author} {\bibfnamefont {B.~D.}\ \bibnamefont
  {Serot}}\ and\ \bibinfo {author} {\bibfnamefont {J.~D.}\ \bibnamefont
  {Walecka}},\ }\href@noop {} {\bibfield  {journal} {\bibinfo  {journal} {Adv.
  Nucl. Phys.}\ }\textbf {\bibinfo {volume} {16}},\ \bibinfo {pages} {1}
  (\bibinfo {year} {1986})}\BibitemShut {NoStop}%
\bibitem [{\citenamefont {Anastasio}\ \emph {et~al.}(1981)\citenamefont
  {Anastasio}, \citenamefont {Celenza},\ and\ \citenamefont
  {Shakin}}]{Anastasio1981_PRC23-2273}%
  \BibitemOpen
  \bibfield  {author} {\bibinfo {author} {\bibfnamefont {M.~R.}\ \bibnamefont
  {Anastasio}}, \bibinfo {author} {\bibfnamefont {L.~S.}\ \bibnamefont
  {Celenza}}, \ and\ \bibinfo {author} {\bibfnamefont {C.~M.}\ \bibnamefont
  {Shakin}},\ }\href@noop {} {\bibfield  {journal} {\bibinfo  {journal} {Phys.
  Rev. C}\ }\textbf {\bibinfo {volume} {23}},\ \bibinfo {pages} {2273}
  (\bibinfo {year} {1981})}\BibitemShut {NoStop}%
\bibitem [{\citenamefont {Poschenrieder}\ and\ \citenamefont
  {Weigel}(1988)}]{Poschenrieder-1988-PRC.38.471}%
  \BibitemOpen
  \bibfield  {author} {\bibinfo {author} {\bibfnamefont {P.}~\bibnamefont
  {Poschenrieder}}\ and\ \bibinfo {author} {\bibfnamefont {M.~K.}\ \bibnamefont
  {Weigel}},\ }\href {\doibase 10.1103/PhysRevC.38.471} {\bibfield  {journal}
  {\bibinfo  {journal} {Physical Review C}\ }\textbf {\bibinfo {volume} {38}},\
  \bibinfo {pages} {471} (\bibinfo {year} {1988})}\BibitemShut {NoStop}%
\bibitem [{\citenamefont {Machleidt}(1989)}]{Machleidt1989_ANP19-189}%
  \BibitemOpen
  \bibfield  {author} {\bibinfo {author} {\bibfnamefont {R.}~\bibnamefont
  {Machleidt}},\ }\href@noop {} {\bibfield  {journal} {\bibinfo  {journal}
  {Adv. Nucl. Phys.}\ }\textbf {\bibinfo {volume} {19}},\ \bibinfo {pages}
  {189} (\bibinfo {year} {1989})}\BibitemShut {NoStop}%
\bibitem [{\citenamefont {Bjorken}\ and\ \citenamefont
  {Drell}(1965)}]{Bjorken1965RQM}%
  \BibitemOpen
  \bibfield  {author} {\bibinfo {author} {\bibfnamefont {J.~D.}\ \bibnamefont
  {Bjorken}}\ and\ \bibinfo {author} {\bibfnamefont {S.~D.}\ \bibnamefont
  {Drell}},\ }\href@noop {} {\emph {\bibinfo {title} {{Relativistic Quantum
  Mechanics}}}},\ International Series In Pure and Applied Physics\ (\bibinfo
  {publisher} {McGraw-Hill},\ \bibinfo {address} {New York},\ \bibinfo {year}
  {1965})\BibitemShut {NoStop}%
\bibitem [{\citenamefont {Oyamatsu}\ \emph {et~al.}(2010)\citenamefont
  {Oyamatsu}, \citenamefont {Iida},\ and\ \citenamefont
  {Koura}}]{Oyamatsu2010PRC}%
  \BibitemOpen
  \bibfield  {author} {\bibinfo {author} {\bibfnamefont {K.}~\bibnamefont
  {Oyamatsu}}, \bibinfo {author} {\bibfnamefont {K.}~\bibnamefont {Iida}}, \
  and\ \bibinfo {author} {\bibfnamefont {H.}~\bibnamefont {Koura}},\ }\href
  {\doibase 10.1103/PhysRevC.82.027301} {\bibfield  {journal} {\bibinfo
  {journal} {Phys. Rev. C}\ }\textbf {\bibinfo {volume} {82}},\ \bibinfo
  {pages} {027301} (\bibinfo {year} {2010})}\BibitemShut {NoStop}%
\bibitem [{\citenamefont {Abrahamyan}\ \emph {et~al.}(2012)\citenamefont
  {Abrahamyan}, \citenamefont {Ahmed}, \citenamefont {Albataineh},
  \citenamefont {Aniol}, \citenamefont {Armstrong}, \citenamefont {Armstrong},
  \citenamefont {Averett}, \citenamefont {Babineau}, \citenamefont {Barbieri},
  \citenamefont {Bellini}, \citenamefont {Beminiwattha}, \citenamefont
  {Benesch}, \citenamefont {Benmokhtar}, \citenamefont {Bielarski},
  \citenamefont {Boeglin}, \citenamefont {Camsonne}, \citenamefont {Canan},
  \citenamefont {Carter}, \citenamefont {Cates}, \citenamefont {Chen},
  \citenamefont {Chen}, \citenamefont {Hen}, \citenamefont {Cusanno},
  \citenamefont {Dalton}, \citenamefont {De~Leo}, \citenamefont {de~Jager},
  \citenamefont {Deconinck}, \citenamefont {Decowski}, \citenamefont {Deng},
  \citenamefont {Deur}, \citenamefont {Dutta}, \citenamefont {Etile},
  \citenamefont {Flay}, \citenamefont {Franklin}, \citenamefont {Friend},
  \citenamefont {Frullani}, \citenamefont {Fuchey}, \citenamefont {Garibaldi},
  \citenamefont {Gasser}, \citenamefont {Gilman}, \citenamefont {Giusa},
  \citenamefont {Glamazdin}, \citenamefont {Gomez}, \citenamefont {Grames},
  \citenamefont {Gu}, \citenamefont {Hansen}, \citenamefont {Hansknecht},
  \citenamefont {Higinbotham}, \citenamefont {Holmes}, \citenamefont
  {Holmstrom}, \citenamefont {Horowitz}, \citenamefont {Hoskins}, \citenamefont
  {Huang}, \citenamefont {Hyde}, \citenamefont {Itard}, \citenamefont {Jen},
  \citenamefont {Jensen}, \citenamefont {Jin}, \citenamefont {Johnston},
  \citenamefont {Kelleher}, \citenamefont {Kliakhandler}, \citenamefont {King},
  \citenamefont {Kowalski}, \citenamefont {Kumar}, \citenamefont {Leacock},
  \citenamefont {Leckey}, \citenamefont {Lee}, \citenamefont {LeRose},
  \citenamefont {Lindgren}, \citenamefont {Liyanage}, \citenamefont {Lubinsky},
  \citenamefont {Mammei}, \citenamefont {Mammoliti}, \citenamefont
  {Margaziotis}, \citenamefont {Markowitz}, \citenamefont {McCreary},
  \citenamefont {McNulty}, \citenamefont {Mercado}, \citenamefont {Meziani},
  \citenamefont {Michaels}, \citenamefont {Mihovilovic}, \citenamefont
  {Muangma}, \citenamefont {Mu\~noz Camacho}, \citenamefont {Nanda},
  \citenamefont {Nelyubin}, \citenamefont {Nuruzzaman}, \citenamefont {Oh},
  \citenamefont {Palmer}, \citenamefont {Parno}, \citenamefont {Paschke},
  \citenamefont {Phillips}, \citenamefont {Poelker}, \citenamefont
  {Pomatsalyuk}, \citenamefont {Posik}, \citenamefont {Puckett}, \citenamefont
  {Quinn}, \citenamefont {Rakhman}, \citenamefont {Reimer}, \citenamefont
  {Riordan}, \citenamefont {Rogan}, \citenamefont {Ron}, \citenamefont {Russo},
  \citenamefont {Saenboonruang}, \citenamefont {Saha}, \citenamefont
  {Sawatzky}, \citenamefont {Shahinyan}, \citenamefont {Silwal}, \citenamefont
  {Sirca}, \citenamefont {Slifer}, \citenamefont {Solvignon}, \citenamefont
  {Souder}, \citenamefont {Sperduto}, \citenamefont {Subedi}, \citenamefont
  {Suleiman}, \citenamefont {Sulkosky}, \citenamefont {Sutera}, \citenamefont
  {Tobias}, \citenamefont {Troth}, \citenamefont {Urciuoli}, \citenamefont
  {Waidyawansa}, \citenamefont {Wang}, \citenamefont {Wexler}, \citenamefont
  {Wilson}, \citenamefont {Wojtsekhowski}, \citenamefont {Yan}, \citenamefont
  {Yao}, \citenamefont {Ye}, \citenamefont {Ye}, \citenamefont {Yim},
  \citenamefont {Zana}, \citenamefont {Zhan}, \citenamefont {Zhang},
  \citenamefont {Zhang}, \citenamefont {Zheng},\ and\ \citenamefont
  {Zhu}}]{Abrahamyan2012PRL}%
  \BibitemOpen
  \bibfield  {author} {\bibinfo {author} {\bibfnamefont {S.}~\bibnamefont
  {Abrahamyan}}, \bibinfo {author} {\bibfnamefont {Z.}~\bibnamefont {Ahmed}},
  \bibinfo {author} {\bibfnamefont {H.}~\bibnamefont {Albataineh}}, \bibinfo
  {author} {\bibfnamefont {K.}~\bibnamefont {Aniol}}, \bibinfo {author}
  {\bibfnamefont {D.~S.}\ \bibnamefont {Armstrong}}, \bibinfo {author}
  {\bibfnamefont {W.}~\bibnamefont {Armstrong}}, \bibinfo {author}
  {\bibfnamefont {T.}~\bibnamefont {Averett}}, \bibinfo {author} {\bibfnamefont
  {B.}~\bibnamefont {Babineau}}, \bibinfo {author} {\bibfnamefont
  {A.}~\bibnamefont {Barbieri}}, \bibinfo {author} {\bibfnamefont
  {V.}~\bibnamefont {Bellini}}, \bibinfo {author} {\bibfnamefont
  {R.}~\bibnamefont {Beminiwattha}}, \bibinfo {author} {\bibfnamefont
  {J.}~\bibnamefont {Benesch}}, \bibinfo {author} {\bibfnamefont
  {F.}~\bibnamefont {Benmokhtar}}, \bibinfo {author} {\bibfnamefont
  {T.}~\bibnamefont {Bielarski}}, \bibinfo {author} {\bibfnamefont
  {W.}~\bibnamefont {Boeglin}}, \bibinfo {author} {\bibfnamefont
  {A.}~\bibnamefont {Camsonne}}, \bibinfo {author} {\bibfnamefont
  {M.}~\bibnamefont {Canan}}, \bibinfo {author} {\bibfnamefont
  {P.}~\bibnamefont {Carter}}, \bibinfo {author} {\bibfnamefont {G.~D.}\
  \bibnamefont {Cates}}, \bibinfo {author} {\bibfnamefont {C.}~\bibnamefont
  {Chen}}, \bibinfo {author} {\bibfnamefont {J.-P.}\ \bibnamefont {Chen}},
  \bibinfo {author} {\bibfnamefont {O.}~\bibnamefont {Hen}}, \bibinfo {author}
  {\bibfnamefont {F.}~\bibnamefont {Cusanno}}, \bibinfo {author} {\bibfnamefont
  {M.~M.}\ \bibnamefont {Dalton}}, \bibinfo {author} {\bibfnamefont
  {R.}~\bibnamefont {De~Leo}}, \bibinfo {author} {\bibfnamefont
  {K.}~\bibnamefont {de~Jager}}, \bibinfo {author} {\bibfnamefont
  {W.}~\bibnamefont {Deconinck}}, \bibinfo {author} {\bibfnamefont
  {P.}~\bibnamefont {Decowski}}, \bibinfo {author} {\bibfnamefont
  {X.}~\bibnamefont {Deng}}, \bibinfo {author} {\bibfnamefont {A.}~\bibnamefont
  {Deur}}, \bibinfo {author} {\bibfnamefont {D.}~\bibnamefont {Dutta}},
  \bibinfo {author} {\bibfnamefont {A.}~\bibnamefont {Etile}}, \bibinfo
  {author} {\bibfnamefont {D.}~\bibnamefont {Flay}}, \bibinfo {author}
  {\bibfnamefont {G.~B.}\ \bibnamefont {Franklin}}, \bibinfo {author}
  {\bibfnamefont {M.}~\bibnamefont {Friend}}, \bibinfo {author} {\bibfnamefont
  {S.}~\bibnamefont {Frullani}}, \bibinfo {author} {\bibfnamefont
  {E.}~\bibnamefont {Fuchey}}, \bibinfo {author} {\bibfnamefont
  {F.}~\bibnamefont {Garibaldi}}, \bibinfo {author} {\bibfnamefont
  {E.}~\bibnamefont {Gasser}}, \bibinfo {author} {\bibfnamefont
  {R.}~\bibnamefont {Gilman}}, \bibinfo {author} {\bibfnamefont
  {A.}~\bibnamefont {Giusa}}, \bibinfo {author} {\bibfnamefont
  {A.}~\bibnamefont {Glamazdin}}, \bibinfo {author} {\bibfnamefont
  {J.}~\bibnamefont {Gomez}}, \bibinfo {author} {\bibfnamefont
  {J.}~\bibnamefont {Grames}}, \bibinfo {author} {\bibfnamefont
  {C.}~\bibnamefont {Gu}}, \bibinfo {author} {\bibfnamefont {O.}~\bibnamefont
  {Hansen}}, \bibinfo {author} {\bibfnamefont {J.}~\bibnamefont {Hansknecht}},
  \bibinfo {author} {\bibfnamefont {D.~W.}\ \bibnamefont {Higinbotham}},
  \bibinfo {author} {\bibfnamefont {R.~S.}\ \bibnamefont {Holmes}}, \bibinfo
  {author} {\bibfnamefont {T.}~\bibnamefont {Holmstrom}}, \bibinfo {author}
  {\bibfnamefont {C.~J.}\ \bibnamefont {Horowitz}}, \bibinfo {author}
  {\bibfnamefont {J.}~\bibnamefont {Hoskins}}, \bibinfo {author} {\bibfnamefont
  {J.}~\bibnamefont {Huang}}, \bibinfo {author} {\bibfnamefont {C.~E.}\
  \bibnamefont {Hyde}}, \bibinfo {author} {\bibfnamefont {F.}~\bibnamefont
  {Itard}}, \bibinfo {author} {\bibfnamefont {C.-M.}\ \bibnamefont {Jen}},
  \bibinfo {author} {\bibfnamefont {E.}~\bibnamefont {Jensen}}, \bibinfo
  {author} {\bibfnamefont {G.}~\bibnamefont {Jin}}, \bibinfo {author}
  {\bibfnamefont {S.}~\bibnamefont {Johnston}}, \bibinfo {author}
  {\bibfnamefont {A.}~\bibnamefont {Kelleher}}, \bibinfo {author}
  {\bibfnamefont {K.}~\bibnamefont {Kliakhandler}}, \bibinfo {author}
  {\bibfnamefont {P.~M.}\ \bibnamefont {King}}, \bibinfo {author}
  {\bibfnamefont {S.}~\bibnamefont {Kowalski}}, \bibinfo {author}
  {\bibfnamefont {K.~S.}\ \bibnamefont {Kumar}}, \bibinfo {author}
  {\bibfnamefont {J.}~\bibnamefont {Leacock}}, \bibinfo {author} {\bibfnamefont
  {J.}~\bibnamefont {Leckey}}, \bibinfo {author} {\bibfnamefont {J.~H.}\
  \bibnamefont {Lee}}, \bibinfo {author} {\bibfnamefont {J.~J.}\ \bibnamefont
  {LeRose}}, \bibinfo {author} {\bibfnamefont {R.}~\bibnamefont {Lindgren}},
  \bibinfo {author} {\bibfnamefont {N.}~\bibnamefont {Liyanage}}, \bibinfo
  {author} {\bibfnamefont {N.}~\bibnamefont {Lubinsky}}, \bibinfo {author}
  {\bibfnamefont {J.}~\bibnamefont {Mammei}}, \bibinfo {author} {\bibfnamefont
  {F.}~\bibnamefont {Mammoliti}}, \bibinfo {author} {\bibfnamefont {D.~J.}\
  \bibnamefont {Margaziotis}}, \bibinfo {author} {\bibfnamefont
  {P.}~\bibnamefont {Markowitz}}, \bibinfo {author} {\bibfnamefont
  {A.}~\bibnamefont {McCreary}}, \bibinfo {author} {\bibfnamefont
  {D.}~\bibnamefont {McNulty}}, \bibinfo {author} {\bibfnamefont
  {L.}~\bibnamefont {Mercado}}, \bibinfo {author} {\bibfnamefont {Z.-E.}\
  \bibnamefont {Meziani}}, \bibinfo {author} {\bibfnamefont {R.~W.}\
  \bibnamefont {Michaels}}, \bibinfo {author} {\bibfnamefont {M.}~\bibnamefont
  {Mihovilovic}}, \bibinfo {author} {\bibfnamefont {N.}~\bibnamefont
  {Muangma}}, \bibinfo {author} {\bibfnamefont {C.}~\bibnamefont {Mu\~noz
  Camacho}}, \bibinfo {author} {\bibfnamefont {S.}~\bibnamefont {Nanda}},
  \bibinfo {author} {\bibfnamefont {V.}~\bibnamefont {Nelyubin}}, \bibinfo
  {author} {\bibfnamefont {N.}~\bibnamefont {Nuruzzaman}}, \bibinfo {author}
  {\bibfnamefont {Y.}~\bibnamefont {Oh}}, \bibinfo {author} {\bibfnamefont
  {A.}~\bibnamefont {Palmer}}, \bibinfo {author} {\bibfnamefont
  {D.}~\bibnamefont {Parno}}, \bibinfo {author} {\bibfnamefont {K.~D.}\
  \bibnamefont {Paschke}}, \bibinfo {author} {\bibfnamefont {S.~K.}\
  \bibnamefont {Phillips}}, \bibinfo {author} {\bibfnamefont {B.}~\bibnamefont
  {Poelker}}, \bibinfo {author} {\bibfnamefont {R.}~\bibnamefont
  {Pomatsalyuk}}, \bibinfo {author} {\bibfnamefont {M.}~\bibnamefont {Posik}},
  \bibinfo {author} {\bibfnamefont {A.~J.~R.}\ \bibnamefont {Puckett}},
  \bibinfo {author} {\bibfnamefont {B.}~\bibnamefont {Quinn}}, \bibinfo
  {author} {\bibfnamefont {A.}~\bibnamefont {Rakhman}}, \bibinfo {author}
  {\bibfnamefont {P.~E.}\ \bibnamefont {Reimer}}, \bibinfo {author}
  {\bibfnamefont {S.}~\bibnamefont {Riordan}}, \bibinfo {author} {\bibfnamefont
  {P.}~\bibnamefont {Rogan}}, \bibinfo {author} {\bibfnamefont
  {G.}~\bibnamefont {Ron}}, \bibinfo {author} {\bibfnamefont {G.}~\bibnamefont
  {Russo}}, \bibinfo {author} {\bibfnamefont {K.}~\bibnamefont
  {Saenboonruang}}, \bibinfo {author} {\bibfnamefont {A.}~\bibnamefont {Saha}},
  \bibinfo {author} {\bibfnamefont {B.}~\bibnamefont {Sawatzky}}, \bibinfo
  {author} {\bibfnamefont {A.}~\bibnamefont {Shahinyan}}, \bibinfo {author}
  {\bibfnamefont {R.}~\bibnamefont {Silwal}}, \bibinfo {author} {\bibfnamefont
  {S.}~\bibnamefont {Sirca}}, \bibinfo {author} {\bibfnamefont
  {K.}~\bibnamefont {Slifer}}, \bibinfo {author} {\bibfnamefont
  {P.}~\bibnamefont {Solvignon}}, \bibinfo {author} {\bibfnamefont {P.~A.}\
  \bibnamefont {Souder}}, \bibinfo {author} {\bibfnamefont {M.~L.}\
  \bibnamefont {Sperduto}}, \bibinfo {author} {\bibfnamefont {R.}~\bibnamefont
  {Subedi}}, \bibinfo {author} {\bibfnamefont {R.}~\bibnamefont {Suleiman}},
  \bibinfo {author} {\bibfnamefont {V.}~\bibnamefont {Sulkosky}}, \bibinfo
  {author} {\bibfnamefont {C.~M.}\ \bibnamefont {Sutera}}, \bibinfo {author}
  {\bibfnamefont {W.~A.}\ \bibnamefont {Tobias}}, \bibinfo {author}
  {\bibfnamefont {W.}~\bibnamefont {Troth}}, \bibinfo {author} {\bibfnamefont
  {G.~M.}\ \bibnamefont {Urciuoli}}, \bibinfo {author} {\bibfnamefont
  {B.}~\bibnamefont {Waidyawansa}}, \bibinfo {author} {\bibfnamefont
  {D.}~\bibnamefont {Wang}}, \bibinfo {author} {\bibfnamefont {J.}~\bibnamefont
  {Wexler}}, \bibinfo {author} {\bibfnamefont {R.}~\bibnamefont {Wilson}},
  \bibinfo {author} {\bibfnamefont {B.}~\bibnamefont {Wojtsekhowski}}, \bibinfo
  {author} {\bibfnamefont {X.}~\bibnamefont {Yan}}, \bibinfo {author}
  {\bibfnamefont {H.}~\bibnamefont {Yao}}, \bibinfo {author} {\bibfnamefont
  {Y.}~\bibnamefont {Ye}}, \bibinfo {author} {\bibfnamefont {Z.}~\bibnamefont
  {Ye}}, \bibinfo {author} {\bibfnamefont {V.}~\bibnamefont {Yim}}, \bibinfo
  {author} {\bibfnamefont {L.}~\bibnamefont {Zana}}, \bibinfo {author}
  {\bibfnamefont {X.}~\bibnamefont {Zhan}}, \bibinfo {author} {\bibfnamefont
  {J.}~\bibnamefont {Zhang}}, \bibinfo {author} {\bibfnamefont
  {Y.}~\bibnamefont {Zhang}}, \bibinfo {author} {\bibfnamefont
  {X.}~\bibnamefont {Zheng}}, \ and\ \bibinfo {author} {\bibfnamefont
  {P.}~\bibnamefont {Zhu}} (\bibinfo {collaboration} {PREX Collaboration}),\
  }\href {\doibase 10.1103/PhysRevLett.108.112502} {\bibfield  {journal}
  {\bibinfo  {journal} {Phys. Rev. Lett.}\ }\textbf {\bibinfo {volume} {108}},\
  \bibinfo {pages} {112502} (\bibinfo {year} {2012})}\BibitemShut {NoStop}%
\bibitem [{\citenamefont {Horowitz}\ \emph {et~al.}(2012)\citenamefont
  {Horowitz}, \citenamefont {Ahmed}, \citenamefont {Jen}, \citenamefont
  {Rakhman}, \citenamefont {Souder}, \citenamefont {Dalton}, \citenamefont
  {Liyanage}, \citenamefont {Paschke}, \citenamefont {Saenboonruang},
  \citenamefont {Silwal}, \citenamefont {Franklin}, \citenamefont {Friend},
  \citenamefont {Quinn}, \citenamefont {Kumar}, \citenamefont {McNulty},
  \citenamefont {Mercado}, \citenamefont {Riordan}, \citenamefont {Wexler},
  \citenamefont {Michaels},\ and\ \citenamefont {Urciuoli}}]{Horowitz2012PRC}%
  \BibitemOpen
  \bibfield  {author} {\bibinfo {author} {\bibfnamefont {C.~J.}\ \bibnamefont
  {Horowitz}}, \bibinfo {author} {\bibfnamefont {Z.}~\bibnamefont {Ahmed}},
  \bibinfo {author} {\bibfnamefont {C.-M.}\ \bibnamefont {Jen}}, \bibinfo
  {author} {\bibfnamefont {A.}~\bibnamefont {Rakhman}}, \bibinfo {author}
  {\bibfnamefont {P.~A.}\ \bibnamefont {Souder}}, \bibinfo {author}
  {\bibfnamefont {M.~M.}\ \bibnamefont {Dalton}}, \bibinfo {author}
  {\bibfnamefont {N.}~\bibnamefont {Liyanage}}, \bibinfo {author}
  {\bibfnamefont {K.~D.}\ \bibnamefont {Paschke}}, \bibinfo {author}
  {\bibfnamefont {K.}~\bibnamefont {Saenboonruang}}, \bibinfo {author}
  {\bibfnamefont {R.}~\bibnamefont {Silwal}}, \bibinfo {author} {\bibfnamefont
  {G.~B.}\ \bibnamefont {Franklin}}, \bibinfo {author} {\bibfnamefont
  {M.}~\bibnamefont {Friend}}, \bibinfo {author} {\bibfnamefont
  {B.}~\bibnamefont {Quinn}}, \bibinfo {author} {\bibfnamefont {K.~S.}\
  \bibnamefont {Kumar}}, \bibinfo {author} {\bibfnamefont {D.}~\bibnamefont
  {McNulty}}, \bibinfo {author} {\bibfnamefont {L.}~\bibnamefont {Mercado}},
  \bibinfo {author} {\bibfnamefont {S.}~\bibnamefont {Riordan}}, \bibinfo
  {author} {\bibfnamefont {J.}~\bibnamefont {Wexler}}, \bibinfo {author}
  {\bibfnamefont {R.~W.}\ \bibnamefont {Michaels}}, \ and\ \bibinfo {author}
  {\bibfnamefont {G.~M.}\ \bibnamefont {Urciuoli}},\ }\href {\doibase
  10.1103/PhysRevC.85.032501} {\bibfield  {journal} {\bibinfo  {journal} {Phys.
  Rev. C}\ }\textbf {\bibinfo {volume} {85}},\ \bibinfo {pages} {032501}
  (\bibinfo {year} {2012})}\BibitemShut {NoStop}%
\bibitem [{\citenamefont {Friedman}(2012)}]{Friedman2012NPA}%
  \BibitemOpen
  \bibfield  {author} {\bibinfo {author} {\bibfnamefont {E.}~\bibnamefont
  {Friedman}},\ }\href {\doibase
  https://doi.org/10.1016/j.nuclphysa.2012.09.007} {\bibfield  {journal}
  {\bibinfo  {journal} {Nuclear Physics A}\ }\textbf {\bibinfo {volume}
  {896}},\ \bibinfo {pages} {46} (\bibinfo {year} {2012})}\BibitemShut
  {NoStop}%
\bibitem [{\citenamefont {Garcia-Recio}\ \emph {et~al.}(1992)\citenamefont
  {Garcia-Recio}, \citenamefont {Nieves},\ and\ \citenamefont
  {Oset}}]{Garcia-Recio1992NPA_547_473}%
  \BibitemOpen
  \bibfield  {author} {\bibinfo {author} {\bibfnamefont {C.}~\bibnamefont
  {Garcia-Recio}}, \bibinfo {author} {\bibfnamefont {J.}~\bibnamefont
  {Nieves}}, \ and\ \bibinfo {author} {\bibfnamefont {E.}~\bibnamefont
  {Oset}},\ }\href {\doibase https://doi.org/10.1016/0375-9474(92)90034-H}
  {\bibfield  {journal} {\bibinfo  {journal} {Nuclear Physics A}\ }\textbf
  {\bibinfo {volume} {547}},\ \bibinfo {pages} {473} (\bibinfo {year}
  {1992})}\BibitemShut {NoStop}%
\bibitem [{\citenamefont {Tsang}\ \emph {et~al.}(2012)\citenamefont {Tsang},
  \citenamefont {Stone}, \citenamefont {Camera}, \citenamefont {Danielewicz},
  \citenamefont {Gandolfi}, \citenamefont {Hebeler}, \citenamefont {Horowitz},
  \citenamefont {Lee}, \citenamefont {Lynch}, \citenamefont {Kohley},
  \citenamefont {Lemmon}, \citenamefont {M\"oller}, \citenamefont {Murakami},
  \citenamefont {Riordan}, \citenamefont {Roca-Maza}, \citenamefont
  {Sammarruca}, \citenamefont {Steiner}, \citenamefont {Vida\~na},\ and\
  \citenamefont {Yennello}}]{Tsang2012PRC}%
  \BibitemOpen
  \bibfield  {author} {\bibinfo {author} {\bibfnamefont {M.~B.}\ \bibnamefont
  {Tsang}}, \bibinfo {author} {\bibfnamefont {J.~R.}\ \bibnamefont {Stone}},
  \bibinfo {author} {\bibfnamefont {F.}~\bibnamefont {Camera}}, \bibinfo
  {author} {\bibfnamefont {P.}~\bibnamefont {Danielewicz}}, \bibinfo {author}
  {\bibfnamefont {S.}~\bibnamefont {Gandolfi}}, \bibinfo {author}
  {\bibfnamefont {K.}~\bibnamefont {Hebeler}}, \bibinfo {author} {\bibfnamefont
  {C.~J.}\ \bibnamefont {Horowitz}}, \bibinfo {author} {\bibfnamefont
  {J.}~\bibnamefont {Lee}}, \bibinfo {author} {\bibfnamefont {W.~G.}\
  \bibnamefont {Lynch}}, \bibinfo {author} {\bibfnamefont {Z.}~\bibnamefont
  {Kohley}}, \bibinfo {author} {\bibfnamefont {R.}~\bibnamefont {Lemmon}},
  \bibinfo {author} {\bibfnamefont {P.}~\bibnamefont {M\"oller}}, \bibinfo
  {author} {\bibfnamefont {T.}~\bibnamefont {Murakami}}, \bibinfo {author}
  {\bibfnamefont {S.}~\bibnamefont {Riordan}}, \bibinfo {author} {\bibfnamefont
  {X.}~\bibnamefont {Roca-Maza}}, \bibinfo {author} {\bibfnamefont
  {F.}~\bibnamefont {Sammarruca}}, \bibinfo {author} {\bibfnamefont {A.~W.}\
  \bibnamefont {Steiner}}, \bibinfo {author} {\bibfnamefont {I.}~\bibnamefont
  {Vida\~na}}, \ and\ \bibinfo {author} {\bibfnamefont {S.~J.}\ \bibnamefont
  {Yennello}},\ }\href {\doibase 10.1103/PhysRevC.86.015803} {\bibfield
  {journal} {\bibinfo  {journal} {Phys. Rev. C}\ }\textbf {\bibinfo {volume}
  {86}},\ \bibinfo {pages} {015803} (\bibinfo {year} {2012})}\BibitemShut
  {NoStop}%
\bibitem [{\citenamefont {Bu}\ \emph {et~al.}(2016)\citenamefont {Bu},
  \citenamefont {Li},\ and\ \citenamefont {Schulze}}]{Bu2016CPL}%
  \BibitemOpen
  \bibfield  {author} {\bibinfo {author} {\bibfnamefont {Q.-Y.}\ \bibnamefont
  {Bu}}, \bibinfo {author} {\bibfnamefont {Z.-H.}\ \bibnamefont {Li}}, \ and\
  \bibinfo {author} {\bibfnamefont {H.-J.}\ \bibnamefont {Schulze}},\ }\href
  {\doibase 10.1088/0256-307X/33/3/032101} {\bibfield  {journal} {\bibinfo
  {journal} {Chinese Physics Letters}\ }\textbf {\bibinfo {volume} {33}},\
  \bibinfo {eid} {32101} (\bibinfo {year} {2016})}\BibitemShut {NoStop}%
\bibitem [{\citenamefont {Yao}\ \emph {et~al.}(2012)\citenamefont {Yao},
  \citenamefont {Baroni}, \citenamefont {Bender},\ and\ \citenamefont
  {Heenen}}]{Yao2012PRC}%
  \BibitemOpen
  \bibfield  {author} {\bibinfo {author} {\bibfnamefont {J.-M.}\ \bibnamefont
  {Yao}}, \bibinfo {author} {\bibfnamefont {S.}~\bibnamefont {Baroni}},
  \bibinfo {author} {\bibfnamefont {M.}~\bibnamefont {Bender}}, \ and\ \bibinfo
  {author} {\bibfnamefont {P.-H.}\ \bibnamefont {Heenen}},\ }\href {\doibase
  10.1103/PhysRevC.86.014310} {\bibfield  {journal} {\bibinfo  {journal} {Phys.
  Rev. C}\ }\textbf {\bibinfo {volume} {86}},\ \bibinfo {pages} {014310}
  (\bibinfo {year} {2012})}\BibitemShut {NoStop}%
\bibitem [{\citenamefont {Alam}\ \emph {et~al.}(2014)\citenamefont {Alam},
  \citenamefont {Agrawal}, \citenamefont {De}, \citenamefont {Samaddar},\ and\
  \citenamefont {Col\`o}}]{Alam2014PRC}%
  \BibitemOpen
  \bibfield  {author} {\bibinfo {author} {\bibfnamefont {N.}~\bibnamefont
  {Alam}}, \bibinfo {author} {\bibfnamefont {B.~K.}\ \bibnamefont {Agrawal}},
  \bibinfo {author} {\bibfnamefont {J.~N.}\ \bibnamefont {De}}, \bibinfo
  {author} {\bibfnamefont {S.~K.}\ \bibnamefont {Samaddar}}, \ and\ \bibinfo
  {author} {\bibfnamefont {G.}~\bibnamefont {Col\`o}},\ }\href {\doibase
  10.1103/PhysRevC.90.054317} {\bibfield  {journal} {\bibinfo  {journal} {Phys.
  Rev. C}\ }\textbf {\bibinfo {volume} {90}},\ \bibinfo {pages} {054317}
  (\bibinfo {year} {2014})}\BibitemShut {NoStop}%
\bibitem [{\citenamefont {Wiringa}\ \emph {et~al.}(1995)\citenamefont
  {Wiringa}, \citenamefont {Stoks},\ and\ \citenamefont
  {Schiavilla}}]{Wiringa-1995-PhysRevC.51.38}%
  \BibitemOpen
  \bibfield  {author} {\bibinfo {author} {\bibfnamefont {R.~B.}\ \bibnamefont
  {Wiringa}}, \bibinfo {author} {\bibfnamefont {V.~G.~J.}\ \bibnamefont
  {Stoks}}, \ and\ \bibinfo {author} {\bibfnamefont {R.}~\bibnamefont
  {Schiavilla}},\ }\href {\doibase 10.1103/PhysRevC.51.38} {\bibfield
  {journal} {\bibinfo  {journal} {Phys. Rev. C}\ }\textbf {\bibinfo {volume}
  {51}},\ \bibinfo {pages} {38} (\bibinfo {year} {1995})}\BibitemShut {NoStop}%
\bibitem [{\citenamefont {Li}\ \emph {et~al.}(2008{\natexlab{b}})\citenamefont
  {Li}, \citenamefont {Lombardo}, \citenamefont {Schulze},\ and\ \citenamefont
  {Zuo}}]{LI-ZH2008_PRC77-034316}%
  \BibitemOpen
  \bibfield  {author} {\bibinfo {author} {\bibfnamefont {Z.~H.}\ \bibnamefont
  {Li}}, \bibinfo {author} {\bibfnamefont {U.}~\bibnamefont {Lombardo}},
  \bibinfo {author} {\bibfnamefont {H.-J.}\ \bibnamefont {Schulze}}, \ and\
  \bibinfo {author} {\bibfnamefont {W.}~\bibnamefont {Zuo}},\ }\href@noop {}
  {\bibfield  {journal} {\bibinfo  {journal} {Phys. Rev. C}\ }\textbf {\bibinfo
  {volume} {77}},\ \bibinfo {pages} {034316} (\bibinfo {year}
  {2008}{\natexlab{b}})}\BibitemShut {NoStop}%
\bibitem [{\citenamefont {Li}\ and\ \citenamefont
  {Schulze}(2008)}]{LI-ZH2008_PRC78-028801}%
  \BibitemOpen
  \bibfield  {author} {\bibinfo {author} {\bibfnamefont {Z.~H.}\ \bibnamefont
  {Li}}\ and\ \bibinfo {author} {\bibfnamefont {H.-J.}\ \bibnamefont
  {Schulze}},\ }\href {\doibase 10.1103/PhysRevC.78.028801} {\bibfield
  {journal} {\bibinfo  {journal} {Phys. Rev. C}\ }\textbf {\bibinfo {volume}
  {78}},\ \bibinfo {pages} {028801} (\bibinfo {year} {2008})}\BibitemShut
  {NoStop}%
\bibitem [{\citenamefont {Friar}\ and\ \citenamefont
  {Negele}(1973)}]{Friar1973NPA}%
  \BibitemOpen
  \bibfield  {author} {\bibinfo {author} {\bibfnamefont {J.}~\bibnamefont
  {Friar}}\ and\ \bibinfo {author} {\bibfnamefont {J.}~\bibnamefont {Negele}},\
  }\href {\doibase https://doi.org/10.1016/0375-9474(73)90039-0} {\bibfield
  {journal} {\bibinfo  {journal} {Nuclear Physics A}\ }\textbf {\bibinfo
  {volume} {212}},\ \bibinfo {pages} {93} (\bibinfo {year} {1973})}\BibitemShut
  {NoStop}%
\bibitem [{\citenamefont {Baker}(1964)}]{Baker1964PR}%
  \BibitemOpen
  \bibfield  {author} {\bibinfo {author} {\bibfnamefont {A.}~\bibnamefont
  {Baker}},\ }\href {\doibase 10.1103/PhysRev.134.B240} {\bibfield  {journal}
  {\bibinfo  {journal} {Phys. Rev.}\ }\textbf {\bibinfo {volume} {134}},\
  \bibinfo {pages} {B240} (\bibinfo {year} {1964})}\BibitemShut {NoStop}%
\bibitem [{\citenamefont {Yennie}\ \emph {et~al.}(1954)\citenamefont {Yennie},
  \citenamefont {Ravenhall},\ and\ \citenamefont {Wilson}}]{Yennie1954PR}%
  \BibitemOpen
  \bibfield  {author} {\bibinfo {author} {\bibfnamefont {D.~R.}\ \bibnamefont
  {Yennie}}, \bibinfo {author} {\bibfnamefont {D.~G.}\ \bibnamefont
  {Ravenhall}}, \ and\ \bibinfo {author} {\bibfnamefont {R.~N.}\ \bibnamefont
  {Wilson}},\ }\href {\doibase 10.1103/PhysRev.95.500} {\bibfield  {journal}
  {\bibinfo  {journal} {Phys. Rev.}\ }\textbf {\bibinfo {volume} {95}},\
  \bibinfo {pages} {500} (\bibinfo {year} {1954})}\BibitemShut {NoStop}%
\bibitem [{\citenamefont {Antonov}\ \emph {et~al.}(2005)\citenamefont
  {Antonov}, \citenamefont {Kadrev}, \citenamefont {Gaidarov}, \citenamefont
  {Guerra}, \citenamefont {Sarriguren}, \citenamefont {Udias}, \citenamefont
  {Lukyanov}, \citenamefont {Zemlyanaya},\ and\ \citenamefont
  {Krumova}}]{Antonov2005PRC}%
  \BibitemOpen
  \bibfield  {author} {\bibinfo {author} {\bibfnamefont {A.~N.}\ \bibnamefont
  {Antonov}}, \bibinfo {author} {\bibfnamefont {D.~N.}\ \bibnamefont {Kadrev}},
  \bibinfo {author} {\bibfnamefont {M.~K.}\ \bibnamefont {Gaidarov}}, \bibinfo
  {author} {\bibfnamefont {E.~M.~d.}\ \bibnamefont {Guerra}}, \bibinfo {author}
  {\bibfnamefont {P.}~\bibnamefont {Sarriguren}}, \bibinfo {author}
  {\bibfnamefont {J.~M.}\ \bibnamefont {Udias}}, \bibinfo {author}
  {\bibfnamefont {V.~K.}\ \bibnamefont {Lukyanov}}, \bibinfo {author}
  {\bibfnamefont {E.~V.}\ \bibnamefont {Zemlyanaya}}, \ and\ \bibinfo {author}
  {\bibfnamefont {G.~Z.}\ \bibnamefont {Krumova}},\ }\href {\doibase
  10.1103/PhysRevC.72.044307} {\bibfield  {journal} {\bibinfo  {journal} {Phys.
  Rev. C}\ }\textbf {\bibinfo {volume} {72}},\ \bibinfo {pages} {044307}
  (\bibinfo {year} {2005})}\BibitemShut {NoStop}%
\bibitem [{\citenamefont {Sarriguren}\ \emph {et~al.}(2007)\citenamefont
  {Sarriguren}, \citenamefont {Gaidarov}, \citenamefont {Guerra},\ and\
  \citenamefont {Antonov}}]{Sarriguren2007PRC}%
  \BibitemOpen
  \bibfield  {author} {\bibinfo {author} {\bibfnamefont {P.}~\bibnamefont
  {Sarriguren}}, \bibinfo {author} {\bibfnamefont {M.~K.}\ \bibnamefont
  {Gaidarov}}, \bibinfo {author} {\bibfnamefont {E.~M.~d.}\ \bibnamefont
  {Guerra}}, \ and\ \bibinfo {author} {\bibfnamefont {A.~N.}\ \bibnamefont
  {Antonov}},\ }\href {\doibase 10.1103/PhysRevC.76.044322} {\bibfield
  {journal} {\bibinfo  {journal} {Phys. Rev. C}\ }\textbf {\bibinfo {volume}
  {76}},\ \bibinfo {pages} {044322} (\bibinfo {year} {2007})}\BibitemShut
  {NoStop}%
\bibitem [{\citenamefont {Roca-Maza}\ \emph {et~al.}(2008)\citenamefont
  {Roca-Maza}, \citenamefont {Centelles}, \citenamefont {Salvat},\ and\
  \citenamefont {Vi\~nas}}]{Roca-Maza2008PRC}%
  \BibitemOpen
  \bibfield  {author} {\bibinfo {author} {\bibfnamefont {X.}~\bibnamefont
  {Roca-Maza}}, \bibinfo {author} {\bibfnamefont {M.}~\bibnamefont
  {Centelles}}, \bibinfo {author} {\bibfnamefont {F.}~\bibnamefont {Salvat}}, \
  and\ \bibinfo {author} {\bibfnamefont {X.}~\bibnamefont {Vi\~nas}},\ }\href
  {\doibase 10.1103/PhysRevC.78.044332} {\bibfield  {journal} {\bibinfo
  {journal} {Phys. Rev. C}\ }\textbf {\bibinfo {volume} {78}},\ \bibinfo
  {pages} {044332} (\bibinfo {year} {2008})}\BibitemShut {NoStop}%
\bibitem [{\citenamefont {Roca-Maza}\ \emph {et~al.}(2013)\citenamefont
  {Roca-Maza}, \citenamefont {Centelles}, \citenamefont {Salvat},\ and\
  \citenamefont {Vi\~nas}}]{Roca-Maza2013PRC}%
  \BibitemOpen
  \bibfield  {author} {\bibinfo {author} {\bibfnamefont {X.}~\bibnamefont
  {Roca-Maza}}, \bibinfo {author} {\bibfnamefont {M.}~\bibnamefont
  {Centelles}}, \bibinfo {author} {\bibfnamefont {F.}~\bibnamefont {Salvat}}, \
  and\ \bibinfo {author} {\bibfnamefont {X.}~\bibnamefont {Vi\~nas}},\ }\href
  {\doibase 10.1103/PhysRevC.87.014304} {\bibfield  {journal} {\bibinfo
  {journal} {Phys. Rev. C}\ }\textbf {\bibinfo {volume} {87}},\ \bibinfo
  {pages} {014304} (\bibinfo {year} {2013})}\BibitemShut {NoStop}%
\bibitem [{\citenamefont {Chu}\ \emph {et~al.}(2009)\citenamefont {Chu},
  \citenamefont {Ren}, \citenamefont {Dong},\ and\ \citenamefont
  {Wang}}]{Chu2009PRC}%
  \BibitemOpen
  \bibfield  {author} {\bibinfo {author} {\bibfnamefont {Y.}~\bibnamefont
  {Chu}}, \bibinfo {author} {\bibfnamefont {Z.}~\bibnamefont {Ren}}, \bibinfo
  {author} {\bibfnamefont {T.}~\bibnamefont {Dong}}, \ and\ \bibinfo {author}
  {\bibfnamefont {Z.~W.}\ \bibnamefont {Wang}},\ }\href {\doibase
  10.1103/PhysRevC.79.044313} {\bibfield  {journal} {\bibinfo  {journal} {Phys.
  Rev. C}\ }\textbf {\bibinfo {volume} {79}},\ \bibinfo {pages} {044313}
  (\bibinfo {year} {2009})}\BibitemShut {NoStop}%
\end{thebibliography}%


\end{document}